\documentclass[letterpaper,usenatbib]{mn2e}
\usepackage{rotating}
\usepackage{epsfig}
\usepackage{mathtools}
\usepackage{times}
\def\gtsim{~\rlap{$>$}{\lower 1.0ex\hbox{$\sim$}}}
\def\ltsim{~\rlap{$<$}{\lower 1.0ex\hbox{$\sim$}}}

\title[PAH and FIR emission from NGC 2403 and M83]
    {The relationship between polycyclic aromatic hydrocarbon emission and far-infrared dust emission from NGC 2403 and M83}

\author[A. G. Jones et al.]
    {A. G. Jones$^1$, G. J. Bendo$^{1,2}$, M. Baes$^3$, M. Boquien$^4$, A. Boselli$^4$, I. De Looze$^3$,\newauthor J. Fritz$^3$, F. Galliano$^5$, T. M. Hughes$^3$, V. Lebouteiller$^5$, N. Lu$^6$, S. C. Madden$^5$, \newauthor A. R\'emy-Ruyer$^{7,5}$, M. W. L. Smith$^8$, L.Spinoglio$^9$, A. A. Zijlstra$^1$\\
    $^1$  Jodrell Bank Centre for Astrophysics, 
          School of Physics and Astronomy, University of Manchester, 
          Oxford Road, Manchester M13 9PL, United Kingdom\\
    $^2$  UK ALMA Regional Centre Node\\
    $^3$  Sterrenkundig Observatorium, Universiteit Gent, 
          Krijgslaan 281 S9,  
          B-9000 Gent, Belgium\\
    $^4$  Aix-Marseille Universit\'e, CNRS, LAM (Laboratoire d'Astrophysique de Marseille) 
          UMR 7326, 13388 Marseille, France\\
    $^5$  Laboratoire AIM, CEA, Universit\'e Paris Diderot, 
          IRFU/Service d'Astrophysique, Bat. 709, 
          91191 Gif-sur-Yvette, France\\
    $^6$  NHSC/IPAC, 100-22 Caltech, Pasadena, CA 91125, USA \\
    $^7$  Institut d'Astrophysique Spatiale, CNRS, UMR8617, 91405, Orsay, France \\
    $^8$  School of Physics and Astronomy, Cardiff University, 
          Queens Buildings, The Parade, Cardiff CF24 3AA, United Kingdom\\
    $^9$  Istituto di Fisica dello Spazio Interplanetario, INAF, 
          Via Fosso del Cavaliere 100, I-00133 Roma, Italy
}
\date{}
\pagerange{\pageref{firstpage}--\pageref{lastpage}}
\pubyear{}

\begin{document}
\label{firstpage}
\maketitle

\begin{abstract}
We examine the relation between polycyclic aromatic hydrocarbon (PAH) emission at 8~$\mu$m  and far-infrared emission from hot dust grains at 24~$\mu$m and from large dust grains at 160 and 250~$\mu$m in the nearby spiral galaxies NGC 2403 and M83 using data from the {\it Spitzer} Space Telescope and {\it Herschel} Space Observatory.  We find that the PAH emission in NGC 2403 is better correlated with emission at 250~$\mu$m from dust heated by the diffuse interstellar radiation field (ISRF) and that the 8/250~$\mu$m surface brightness ratio is well-correlated with the stellar surface brightness as measured at 3.6~$\mu$m.  This implies that the PAHs in NGC 2403 are intermixed with cold large dust grains in the diffuse interstellar medium (ISM) and that the PAHs are excited by the diffuse ISRF.  In M83, the PAH emission appears more strongly correlated with 160~$\mu$m emission originating from large dust grains heated by star forming regions.  However, the PAH emission in M83 is low where the 24~$\mu$m emission peaks within star forming regions, and enhancements in the 8/160~$\mu$m surface brightness ratios appear offset relative to the dust and the star forming regions within the spiral arms.  This suggests that the PAHs observed in the 8~$\mu$m band are not excited locally within star forming regions but either by light escaping non-axisymmetrically from star forming regions or locally by young, non-photoionising stars that have migrated downstream from the spiral density waves.  The results from just these two galaxies show that PAHs may be excited by different stellar populations in different spiral galaxies.
\end{abstract}


\section{Introduction}
\label{s_intro}
\addtocounter{footnote}{9}

Polycyclic aromatic hydrocarbons (PAHs) are large carbon molecules that can be thought of as the transition between the gaseous and solid phases of the interstellar medium (ISM).  They are commonly identified as the source of emission for multiple broad emission features in the mid-infrared, including those at 3.3, 6.2, 7.7, 11.3, 12.7~$\mu$m \citep{2008tielens}.  While these complex molecules can be excited by optical and infrared photons \citep{2002li}, excitation by ultraviolet photons is more efficient \citep{2008tielens}.

Many authors have sought to calibrate PAH emission as an extragalactic star formation tracer, mainly because it is relatively unaffected by dust extinction and because PAHs emit at shorter wavelengths than hot dust and can therefore be imaged with better angular resolutions.  While early analyses with the Infrared Space Observatory \citep{1996kessler} found some evidence for a relation between PAH emission and other star formation tracers \citep{2001roussel,2004forster}, later work, including work with the {\it Spitzer} Space Telescope \citep{2004werner}, demonstrated that the PAH emission was actually poorly correlated with other star formation tracers.  PAH emission appeared suppressed relative to other star formation tracers in star forming regions or overluminous in diffuse regions \citep{2004boselli,2004helou,2005calzetti,2007prescott,2008bendo}.  The ratio of PAH to hot dust continuum emission was also found to decrease as metallicity decreased \citep{2005engelbracht,2006madden,2007calzetti,2008engelbracht,2008galliano,2008gordon,2013galametz}. The common explanations are either that low metallicity regions contain fewer PAHs or that the PAHs are exposed to harder ultraviolet radiation in low-metallicity environments with less dust attenuation.  None the less, globally-integrated PAH emission has been shown to be correlated with other globally-integrated star formation tracers, and methods have been developed for calculating extinction-corrected star formation rates using a combination of H$\alpha$ and PAH emission in the 8~$\mu$m {\it Spitzer} band \citep{2008zhu,2009kennicutt}, although \citet{2007calzetti} warns that even global measurements could be affected by metallicity effects.

In contrast, a few authors have found that PAH emission was associated with emission at $>100~\mu$m that will primarily trace $\ltsim30$~K large dust grains.  One of the first groups to draw attention to this was \citet{2002haas}, who demonstrated that PAH emission was better correlated with 850~$\mu$m from cold dust than 15~$\mu$m emission from hot dust, although this analysis was limited to regions with high infrared surface brightnesses.  The ISO-based analysis by \citet{2004boselli} also implied that PAHs were associated with diffuse dust rather than star forming regions.  Later work by \citet{2006bendo} and \cite{2008bendo} demonstrated that the PAH emission showed a strong correlation with 160~$\mu$m emission and that the 8/160~$\mu$m surface brightness ratio was dependent upon the 160~$\mu$m surface brightness.  These results combined with the breakdown in the relation between the PAH and hot dust emission implied that the PAHs were primarily associated with dust in the diffuse ISM or in cold molecular clouds near star forming regions and that both the PAHs and the large dust grains were heated by the same radiation field.

Data from the {\it Herschel} Space Observatory \citep{2010pilbratt} can be used to further study the relation between PAHs and large dust grains.  The telescope is able to resolve emission at 160 and 250~$\mu$m on $<18$~arcsec scales, which is a marked improvement in comparison to the 38~arcsec scales that could be resolved by {\it Spitzer} at 160~$\mu$m.  Moreover, \citet{2010bendo}, \citet{2011boquien}, and \citet{2012bendo} have demonstrated that $\leq160$ and $\geq250$~$\mu$m emission from nearby galaxies may originate from dust heated by different sources.  The 70/160 and 160/250~$\mu$m surface brightness ratios were typically correlated with star formation tracers such as ultraviolet, H$\alpha$, and 24~$\mu$m emission and peaked in locations with strong star formation, suggesting that the dust seen at $\leq160$~$\mu$m is primarily heated locally by star forming regions.  Meanwhile, the 250/350 and 350/500~$\mu$m ratios were more strongly correlated with near-infrared emission and generally varied radially in the same way as the emission from the total stellar population (including both young, intermediate-aged, and evolved stars), demosntrating that the dust was primarily heated by the diffuse interstellar radiation field (ISRF) from these stars.  PAH emission can be compared to dust emission observed by {\it Herschel} to determine which of these two dust components are more closely associated with PAH emission, which would ultimately lead to a better understanding of how the PAHs are excited and how they survive in certain environments in the ISM.

So far, the relation between PAH and dust emission has been investigated using {\it Herschel} data for only two galaxies.  \cite{2014calapa} have shown that 8~$\mu$m emission from PAHs in M33 is well correlated with 250~$\mu$m emission.  They go on to further demonstrate the 8/250~$\mu$m ratio is correlated with the 3.6~$\mu$m band tracing the total stellar population, implying that the PAHs are excited by the diffuse ISRF.  \citet{2014lu} present an alternative analysis with M81 in which they divide the PAH emission into components heated by two sources: a component heated by star forming regions traced by H$\alpha$ emission and a component heated by the diffuse ISRF traced by the cold dust emission at 500~$\mu$m emission.  The results show that most ($\sim85$\%) of the 8~$\mu$m emission from diffuse regions is associated with the cold dust emission, while in star forming regions, most ($\sim60$\%) of the 8~$\mu$m emission is excited by young stars.

The goal of this paper, which is a continuation of the work by \citet{2013jones}, is to further study the relationship between PAH emission at 8~$\mu$m and far-infrared emission from large dust grains using {\it Herschel} Space Observatory \citep{2010pilbratt} far-infrared images of NGC 2403 and M83.  These are two of the fourteen nearby galaxies within the Very Nearby Galaxies Survey (VNGS; PI: C. Wilson), a {\it Herschel}-SPIRE Local Galaxies Guaranteed Time Program.  The VNGS was meant to sample galaxies with multiple morphological and active galactic nucleus types, and includes several well-studied galaxies including the Antennae Galaxies, Arp 220, Centaurus A, M51, and NGC 1068.  These two specific galaxies were selected because they are non-interacting nearby ($<10$~Mpc) spiral galaxies with an inclination from face on $\leq~60^{o}$ and major axes $>10$~arcmin\footnote{M81 is also in the VNGS, but the analysis of PAH emission from that galaxy is covered by \citet{2014lu}.}  The basic properties of these galaxies are given in Table \ref{t_galaxies}.

NGC 2403 is an SAB(s)cd galaxy \citep{1991devaucouleurs} with no clear bulge and flocculent spiral structure \citep{1987elmegreen}.  Since the brightest star forming regions are found well outside the centre of the galaxy, it is easy to differentiate between effects related to star forming regions and either effects related to the evolved stellar population (which peaks in the centre of the galaxy) or effects tied to galactocentric radius.  This has been exploited previously to illustrate how PAH emission is inhibited relative to hot dust emission in star forming regions \citep{2008bendo} and to differentiate between different heating sources for the dust seen at 70-500~$\mu$m \citep{2012bendo}.  M83 (NGC 5236) is an SAB(s)c galaxy \citep{1991devaucouleurs} with a bright starburst nucleus \citep{1983bohlin} and two strongly defined grand design spiral arms \citep{1998elmegreen}.  Since we can resolve the spiral structure with {\it Herschel}, we can compare the properties of arm and interarm regions quite effectively.  Both galaxies are at similar distances; we can resolve structures of $<400$~pc in the {\it Herschel} data.  Although both of these galaxies are late-type spiral galaxies, they have the potential to yield different information on how PAHs relate to the far-infrared emission from large dust grains.

We focus our analysis on the {\it Spitzer} 24~$\mu$m data, which trace emission from very small grains and hot dust heated locally by star forming regions, and {\it Herschel} 160 and 250~$\mu$m data, which trace emission from large dust grains.  The prior analysis by \citet{2008bendo} had shown an association between the 8~$\mu$m and 160~$\mu$m emission, but as stated above, the 160~$\mu$m band may contain significant emission from large dust grains heated by star forming regions, while the 250~$\mu$m band, at least for NGC~2403 and M83, originates more from dust heated by the diffuse ISRF \citep{2012bendo} and could be better associated with PAH emission if PAHs are destroyed in star forming regions.  The next shortest waveband for which we have data for these two galaxies is at 70~$\mu$m, but the available 70~$\mu$m data have a lower signal to noise ratio, and the {\it Spitzer} data are strongly affected by latent image artefacts.  Moreover, the 70~$\mu$m emission may include emission from the same sources as the 24~$\mu$m band.  The available 350 and 500~$\mu$m data trace the same thermal component of dust seen at 250~$\mu$m, but because the resolution of those data are coarser compared to the 250~$\mu$m waveband, using the data would provide no additional benefit.

For this analysis, we use the techniques developed by \citet{2008bendo} and \citet{2012bendo} based upon qualitative analyses of surface brightness ratio maps based on images with matching point spread functions (PSFs) and quantitative analyses of the surface brightnesses and surface brightness rations measured in rebinned versions of these images.  Section \ref{s_data} introduces the data and the data preparation steps.  We then present the analysis in Section~\ref{s_analysis_ratios} and then use these results to identify the PAH excitation sources in Section~\ref{s_pahexcitation}.  Following this, we discuss the implications of these results in Section~\ref{s_discussion} and provide a summary in Section \ref{s_conclusions}.

\begin{table*}
\centering
\begin{minipage}{94mm}
\caption{Properties of the sample galaxies$^a$.}
\label{t_galaxies}
\begin{tabular}{@{}lccccc@{}}
\hline
Name 	&
    RA & 
    Dec &
    Hubble &
    Distance &
    Size of Optical \\ 
&
    (J2000) & 
    (J2000) &
    Type &
    (Mpc)$^b$ &
    Disc (arcmin) \\ 
\hline
NGC 2403 & 
    07 36 54.5 &
   +65 35 58 &
   SAB(s)cd &
   3.2 $\pm$0.3 &
   $22.0 \times 12.3$ \\
M83 &
    13 37 00.3 &
    -29 52 04 &
    SAB(s)c &
    4.5$\pm$ 0.2 &
    $12.9 \times 11.3$ \\
\hline
\end{tabular}
$^{a}$ Data are taken from \cite{1991devaucouleurs} unless otherwise specified.\\
$^{b}$ Distances are taken from \cite{2001freedman}.\\ 
\end{minipage}
\end{table*}

\section{Data}
\label{s_data}

The 3.6, 4.5, 5.8 and 8.0~$\mu$m data for NGC~2403 were observed with the Infrared Array Camera \citep[IRAC; ][]{2004fazio} on {\it Spitzer} as part of the {\it Spitzer} Infrared Nearby Galaxies Survey \citep[SINGS; ][]{2003kennicutt}, and the 3.6-8.0~$\mu$m images for M83 were observed with IRAC by the Local Volume Legacy (LVL) Survey \citep{2009dale}.  Both groups used similar drizzle techniques to mosaic basic calibrated data to produce final images with 0.75~arcsec pixels. The full-width at half maxima (FWHMs) of the PSFs are listed in Table~\ref{t_IRAC}.  We also applied correction factors that optimise the data for photometry of extended source emission as suggested by the IRAC Instrument Handbook; these correction factors are listed in Table~\ref{t_IRAC}.  The calibration uncertainty of the data is 3\% \citep{2013irac}.

\begin{table}
\caption{Properties of the IRAC instrument$^a$.}
\label{t_IRAC}
\begin{center}
\begin{tabular}{lcc}
\hline
Channel &
  FWHM  &
  Correction Factors$^b$ \\
\hline
3.6~$\mu$m &
  1.7 &
  0.91 \\
4.5~$\mu$m &
  1.7 &
  0.94 \\
5.8~$\mu$m &
  1.9 &
  0.66 \\
8.0~$\mu$m &
  2.0  &
  0.74 \\
\hline
\end{tabular}
\end{center}
$^a$ These data are given by the IRAC Instrument Handbook \citep{2013irac}\footnote{http://irsa.ipac.caltech.edu/data/SPITZER/docs/irac\\ /iracinstrumenthandbook/IRAC\_Instrument\_Handbook.pdf}.\\
$^b$ These correction factors are for extended, diffuse emission. 
\end{table}

The 24~$\mu$m data were acquired using the Multiband Imaging Photometer for Spitzer \citep[MIPS; ][]{2004rieke} on {\it Spitzer} and were reprocessed by \citet{2012bendomips} using the MIPS Data Analysis Tools \citep{2005gordon} along with multiple modifications.  The final images have pixel scales of 1.5~arcsec, PSF with FWHM of 6~arcsec \citep{2007engelbracht}, and calibration uncertainties of 4\% \citep{2007engelbracht}.

The 160~$\mu$m data are updated versions of the 160~$\mu$m data published by \citet{2012bendo} and \citet{2012foyle}.  The galaxies were observed at 160~$\mu$m with the Photodetector Array Camera and Spectrometer \citep[PACS; ][]{2010poglitsch} on {\it Herschel} in four pairs of orthogonal scans performed at the 20 arcsec s$^{-1}$ rate.  The observations of NGC~2403 covered a $40\times40$ arcmin region, while the observations of M83 covered a $25\times25$ arcmin region.  The data were processed using the {\it Herschel} Interactive Processing Environment \citep[\small{HIPE}; ][]{2010ott} version 11.1.  We used the standard data processing pipeline, which includes cosmic ray removal and cross-talk corrections, for the individual data frames.  We then remapped the data using {\small SCANAMORPHOS} version 23 \citep{2013roussel}, which also removes additional noise in the data and drift in the background signal.  We applied a colour correction of $1.01 \pm
0.07$, which has a mean value and uncertainty appropriate for emission from a modified blackbody with a temperature between 15 and 40~K and an emissivity function that scales as $\lambda^{-\beta}$ where $\beta$ is between 1 and 2 \citep{2011muller}\footnote{http://herschel.esac.esa.int/twiki/pub/Public/PacsCalibrationWeb\\ /cc\_report\_v1.pdf}.  The FWHM of the PSF is $\sim12$~arcsec \citep{2012lutz}\footnote{
  https://herschel.esac.esa.int/twiki/pub/Public/PacsCalibrationWeb\\ /bolopsf\_20.pdf}, and the flux calibration uncertainty is 5\%
\citep{2013altieri}\footnote{http://herschel.esac.esa.int/Docs/PACS/pdf/pacs\_om.pdf}.

The 250~$\mu$m images, produced using data from the Spectral and Photometric Imaging REceiver \citep[SPIRE; ][]{2010griffin} on {\it Herschel}, are also updated versions of the 250~$\mu$m images originally published by \citet{2012bendo} and \citet{2012foyle}.  The observations consisted of one pair of orthogonal scans using the 30 arcsec s$^{-1}$ scan rate and nominal bias voltage settings.  The maps cover a $30\times30$ arcmin region around NGC~2403 and $40\times40$~arcmin region around M83.  The data were reprocessed using HIPE version 12.1 through a pipeline that includes the standard signal jump correction, cosmic ray removal, low pass filter correction, and bolometer time response corrections, but we used the BRIght Galaxy ADaptive Element method \citep[][Smith et al., in preparation]{2012smith, 2013auld} to remove drift in the background signal and to destripe the data.  The final maps were produced using the naive mapmaker in HIPE and have pixel scales of 6~arcsec.  The FWHM of the PSF is specified by the SPIRE Handbook \citep{2014valtchanov}\footnote{herschel.esac.esa.int/Docs/SPIRE/spire\_handbook.pdf} as 18.1~arcsec, and the calibration uncertainty is 4\% \citep{2013bendo}.  To optimise the data for extended source photometry, we multiplied the data by the point source to extended source conversion factor of 91.289 MJy sr$^{-1}$ (Jy beam$^{-1}$)$^{-1}$ \citep{2014valtchanov} and then applied a colour correction of $0.997 \pm 0.029$, which should be appropriate for a modified blackbody with a temperature between 10 and 40~K and a $\beta$ between 1.5 and 2 \citep{2014valtchanov}.

For a discussion on the spiral density waves in M83 in Section~\ref{s_m83pah}, we also included 0.23~$\mu$m data from the Galaxy Evolution Explorer \citep[GALEX; ][]{2005martin} produced by \citet[][ see also \citealt{2011lee}]{2009dale}.  The images have pixel scales of 1.5~arcsec, PSF FWHM of $\sim6$~arcsec \citep{2005martin}, and calibration uncertainties of $<1$\% \citep{2007morrissey}.  We applied a foreground extinction correction based on $A_{0.23\mu\text{m}}=0.56$ given by \citet{2011lee} based on the \citet{1989cardelli} extinction law with $R_V=3.1$.

For measuring quantitative star formation rates (so that we could identify locations that are strongly influenced by star formation using quantitative criteria), we included H$\alpha$ data for these two galaxies in our analysis.  The H$\alpha$ image for NGC~2403 was originally produced by \citet{2002boselli} using data from the 1.20~m Newton Telescope at the Observatoire de Haute Provence.  The H$\alpha$ image for M83 was produced by \citet{2006meurer} using observations from the Cerro Tololo 1.5 Meter Telescope taken as part of the Survey for Ionization in Neutral Gas Galaxies.  We applied extinction corrections for dust attenuation within the Milky Way using calculations performed by the NASA/IPAC Extragalactic Database\footnote{http://ned.ipac.caltech.edu/} based on data from \citet{1998schlegel}, and we also use data from the literature to correct for [N{\small II}] emission falling within the wavebands covered by the H$\alpha$ filters.  Details on the data are given in Table~\ref{t_hadata}.

\begin{table}
\caption{Properties of and corrections for the H$\alpha$ images}
\label{t_hadata}
\begin{center}
\begin{tabular}{lp{1.8cm}p{1.8cm}}
\hline
Galaxy &                              NGC~2403 &                        M83 \\ \hline
Source &                              \citet{2002boselli} &             \citet{2006meurer} \\
Pixel Scale  (arcsec ~pixel$^{-1}$) & 0.69 &                            0.43 \\
PSF FWHM (arcsec) &                   3 &                               1.6 \\
Calibration uncertainty &             5\% &                             4\% \\
Forground extinction ($A_R$) &        0.87 &                            0.144 \\ \relax
[N{\small II}] / H$\alpha$ ratio &      $0.28 \pm 0.05^a$ &               $0.40 \pm 0.13^b$\\
\hline
\end{tabular}
\end{center}
$^a$ Both the 6548 and 6583 {\AA} [N{\small II}] lines fall within the band covered by the H$\alpha$ filter used in the NGC~2403 observations.  This number represents the ratio of emission from both lines to H$\alpha$ emission measured in the radial strip data from \citet{2010moustakas}.\\
$^b$ The H$\alpha$ filter used in the M83 observations includes emission from only the [N{\small II}] 6583 {\AA} line.  This ratio is based on the ratio of ony that line to H$\alpha$ emission and is calculated using data from \citet{2005boissier}.
\end{table}

\subsection{Data preparation}
\label{s_data_prep}

To study the relation of PAH emission to far-infrared emission from large dust grains, we perform analyses using maps in which the PSFs have been matched to the PSF of the 250~$\mu$m data (with a FWHM of 18~arcsec), and we plot data from images with matching PSFs that have been resampled into 18~arcsec bins that represent individual resolution elements within the maps.  The data from these bins should be statistically independent.  See \citet{2008bendo} and \citet{2012bendo} for additional discussion on this topic.

In the first step of preparing the data, foreground stars were identified by eye and removed from the H$\alpha$, 3.6, 4.5, 5.8, 8 and 24~$\mu$m data; these were typically sources that appeared unresolved and that had 3.6/24~$\mu$m flux density ratios $\gtsim$10.  Next, the data were convolved with kernels from \citet{2011aniano}\footnote{Available from http://www.astro.princeton.edu/$\sim$ganiano/Kernels.html .  Note that the kernels are created using circularised versions of instrumental PSFs.  In the case of the IRAC data, the circularised PSFs have FWHM ranging from 1.9 to 2.8~arcsec, which is larger than the original PSFs.} to match the PSFs of the H$\alpha$, 3.6, 4.5, 5.8, 8, 24 and 160~$\mu$m data to the 18~arcsec PSF of the 250~$\mu$m data. This was done to preserve the colour variations across the data when it was rebinned, and it eliminated the need to perform additional aperture corrections.  The median background was then measured outside of the optical disc of each galaxy in each waveband and subtracted from the data.  For the qualitative map-based analyses, the 3.6 and 8 $\mu$m images were shifted to match the world coordinate systems of the 24, 160 and 250~$\mu$m maps so that we could create 8/24, 8/160, and 8/250~$\mu$m surface brightness ratio maps; the pixel size of each ratio map is set to the pixel size of the image for the longer-wavelength data used in the ratios.  For the analyses on binned data and for producing the profiles in Section~\ref{s_m83pah}, the images were all shifted to match the world coordinate system of the 250~$\mu$m data and then rebinned into 18~arcsec pixels to match the size of the PSF of the 250~$\mu$m data.  The rebinning was done so that the centre of each galaxy was located at the centre of an 18~arcsec bin.

The emission observed in the 4.5-24~$\mu$m bands contains stellar continuum emission.  We remove this stellar emission by subtracting a rescaled version of the IRAC 3.6~$\mu$m image \citep[e.g. ]{2004helou, 2010marble, 2014ciesla}.  The IRAC 3.6~$\mu$m band is suitable for this step because it generally contains unobscured stellar emission \citep{2003lu}.  While hot dust emission may produce 3.6~$\mu$m emission \citep{2009mentuch, 2010mentuch} and while emission from PAHs at 3.3~$\mu$m also falls within the IRAC 3.6~$\mu$m band, the comparison of 3.6~$\mu$m emission to H-band emission by Bendo et al. (2014, submitted) suggests that, on the spatial scales of our data, local enhancements in hot dust and 3.3~$\mu$m PAH emission have a very minor effect on the total 3.6~$\mu$m emission.  The continuum subtraction equations derived by \citet{2004helou} were based on using an earlier version of Starburst99 \citep{1999leitherer} to simulate the infrared stellar SED of a stellar population with a Salpeter initial mass function \citep[IMF; ][]{1955salpeter} and two different metallicities.  From this analysis, Helou et al. derived mean 3.6/8 and 3.6/24~$\mu$m stellar surface brightness ratios that could be used to rescale the 3.6~$\mu$m emission and subtract it from the 8, 4.5, 5.8 and 24~$\mu$m data.  We re-derived these values using a newer version of Starburst99 (version 6.0.3) to simulate a solar metallicity stellar population with a Kroupa IMF \citep{2001kroupa}, which is now becoming more popular to use than the Salpeter IMF.  We also examined the differences resulting from using both the Geneva and Padova stellar evolutionary tracks and found that the selection of one set of tracks over the other did not significantly affect the results.  From these tests, we derive the following equations to subtract the stellar continuum from the 4.5-24~$\mu$m data:
\begin{equation}
\begin{multlined}
I_{\nu}(4.5 \mu \mbox{m}~\mbox{(SCS)} ) \\
  = I_\nu(4.5 \mu \mbox{m}) - (0.60 \pm 0.02) I_\nu(3.6 \mu \mbox{m}) 
\end{multlined}
\end{equation}

\begin{equation}
\begin{multlined}
I_{\nu}(5.8 \mu \mbox{m}~\mbox{(SCS)} ) \\
  = I_\nu(5.8 \mu \mbox{m}) - (0.40 \pm 0.03) I_\nu(3.6 \mu \mbox{m}) 
\end{multlined}
\end{equation}

\begin{equation}
\begin{multlined}
I_{\nu}(8 \mu \mbox{m}~\mbox{(SCS)} ) \\
  = I_\nu(8 \mu \mbox{m}) - (0.246 \pm 0.015) I_\nu(3.6 \mu \mbox{m}) 
\end{multlined}
\end{equation}

\begin{equation}
\begin{multlined}
I_\nu(24 \mu \mbox{m} ~\mbox{(SCS)} ) \\
  = I_\nu(24 \mu \mbox{m}) - (0.033 \pm 0.003) I_\nu(3.6 \mu \mbox{m}). 
\end{multlined}
\end{equation}
In these equations, "SCS" stands for stellar continuum subtracted.  The scaling terms are based on calculations performed at time intervals equally spaced in logarithm space between $10^{7}$ and $10^{10}$ yr.  The values of the scaling terms are based on the mean of the results from using the Geneva and Padova tracks.  The uncertainties are the greater of either the difference in the mean values measured between the results for the two tracks or the larger of the standard deviations measured in the scaling terms derived for the separate tracks. Changing the metallicities to $Z=0.008$ changed the factors by $\ltsim1\sigma$.  The uncertainties in the coefficients translate to a $\ltsim1$\% uncertainty in the corrected 8 and 24~$\mu$m maps, which is negligible compared to the calibration uncertainties.  The 4.5-8.0~$\mu$m coefficients derived here are typically within $1\sigma$ of equivalent coefficients derived in other studies \citep[e.g.][]{2004helou, 2010marble, 2014ciesla}.  The coefficients for the 24~$\mu$m data may disagree with coefficients from other papers by up to 0.012 or $4\sigma$, although the values derived in these other papers differ among each other by 0.018.  However, this correction is so small for the 24~$\mu$m data (typically $\sim1$\% in NGC~2403 and M83) that the relatively high disagreement among the values should not have a major impact on our analysis or on other analyses relying upon this type of stellar continuum subtraction.

The 8~$\mu$m band still contains continuum emission from very hot grains.  In most solar-metallicity galaxies, this continuum emission may constitute $\sim20$\% of the total stellar-continuum-subtracted 8~$\mu$m emission \citep[e.g.][]{2007smith}, although in locations with very weak PAH emission, such as star-forming regions or metal-poor dwarf galaxies, a much higher percentage of the 8~$\mu$m emission may be thermal continuum emission \citep[e.g.][]{2005engelbracht, 2006cannon, 2008engelbracht, 2008gordon}.   To remove the excess dust continuum emission, we use the following equation derived in an empirical analysis of photometric and spectroscopic data by \citet{2010marble}\footnote{The equation given by \citet{2010marble} also includes a term that integrates the emission in frequency and converts the data into units of erg s$^{-1}$ cm$^{-2}$.  Since we are comparing the PAH emission in the 8~$\mu$m band to continuum emission in other bands that is measured in Jy arcsec$^{-2}$, it is easier to keep the 8~$\mu$m data in units of Jy arcsec$^{-2}$, so we do not include the unit conversion term in this equation.}:
\begin{equation}
\label{e_pahdustsub}
\begin{multlined}
I_{\nu}(8 \mu \mbox{m}~{\mbox{(PAH)})} =  (I_{\nu}(8 \mu \mbox{m}~{\mbox{(SCS)})}\\
  - (0.091 + .314 I_{\nu}(8 \mu \mbox{m})/I_{\nu}(24 \mu \mbox{m}))  \\
  \times (I_{\nu}(4.5 \mu \mbox{m}~{\mbox{(SCS)})} + I_{\nu}(5.8 \mu \mbox{m}~{\mbox{(SCS)})})^{0.718} \\
  \times I_{\nu}(24 \mu \mbox{m}~{\mbox{(SCS)})}^{0.282}) 
\end{multlined}
\end{equation}
When this equation is applied to our data, the 8~$\mu$m surface brightnesses typically decrease by $15-20$\%.  Based on the analysis from \citet{2010marble}, the percentage difference between the 8~$\mu$m PAH fluxes calculated using this equation and the fluxes of the spectral features measured spectroscopically is 6\%. Throughout the rest of this paper, when we refer to 8~$\mu$m emission, we are referring to the 8~$\mu$m emission calculated using Equation~\ref{e_pahdustsub}.

For the binned analysis, we wanted to illustrate which bins were more strongly influenced by emission from star forming regions and which regions tend to trace emission from dust predominantly heated by evolved stars. To do this, we created specific star formation rate (SSFR) maps.  We first applied an intrinsic extinction correction to the H$\alpha$ intensities (measured in erg cm$^{-2}$ s$^{-1}$ arcsec$^{-2}$) using 
\begin{equation}
\label{e_hacorr}
\begin{multlined}
I(\mbox{H$\alpha$ (corrected)}) = I(\mbox{H$\alpha$ (observed)})\\
	+ 2.0 \times 10^{-25} (12.5~\mbox{THz})I_{\nu}(24 \mu \mbox{m}) \left( \frac{\mbox{erg cm$^{-2}$ s$^{-1}$}}{\mbox{Jy}} \right),
\end{multlined}
\end{equation}
which is a variant of the correction equation given by \citet{2009kennicutt}.  Since 24~$\mu$m emission has been shown to be associated with H$\alpha$ emission and other star formation tracers \citep[e.g.][]{2005calzetti, 2007calzetti, 2007prescott, 2014bendo}, it is the best band to use when correcting H$\alpha$ emission for intrinsic dust extinction.  The 24~$\mu$m band may also contain emission from diffuse dust heated by the radiation field from evolved stars \citep{2009kennicutt}, which we would expect to affect low surface brightness regions in these galaxies, so low star formation rates derived using Equation~\ref{e_hacorr} should be treated cautiously.

After converting the corrected H$\alpha$ intensities to units of erg s$^{-1}$ pc$^{-2}$ (written as $L(\mbox{H}\alpha)/A$, with $A$ representing the area per pixel in pc$^2$), we used 
\begin{equation}
\Sigma(\mbox{SFR}) = 7.9 \times 10^{-42}\left(\frac{L(\mbox{H}\alpha)}{A}\right)\left(\frac{\mbox{erg s$^{-1}$}}{\mbox{M$_\odot$ yr$^{-1}$}}\right)
\end{equation}
from \citet{1998kennicutt} to calculate star formation rate surface densities $\Sigma(\mbox{SFR})$.  To produce maps of the total stellar surface mass density $\Sigma(\mbox{M$_{\star}$})$, we used
\begin{equation}
\label{e_totalstellar}
\begin{multlined}
\Sigma(\mbox{M$_{\star}$})
   =10^{5.65}
   \left(\frac{I_{\nu}(3.6 \mu \mbox{m})^{2.85}I_{\nu}(4.5 \mu \mbox{m})^{-1.85}\Omega}
   {A}\right)\\  
   \left(\frac{D}{0.05}\right)^{2}
   \left(\frac{\mbox{M$_{\odot}$ arcsec$^{2}$}}
   {\mbox{Jy Mpc$^2$ pc$^2$}}\right)
\end{multlined}
\end{equation}
based on the equation from \citet{2012eskew}.  In this equation, $\Omega$ is the angular area of the bin in the map, and $D$ distance to the source.  We then divided $\Sigma(\mbox{SFR})$ by $\Sigma(\mbox{M$_{\star}$})$ to calculate the SSFR.

\section{Analysis of 8/24, 8/160, and 8/250~$\mu$\lowercase{m} ratios}
\label{s_analysis_ratios}

\subsection{Map-based analysis}
\label{s_analysis_maps}

\begin{figure*}
\begin{center}
\epsfig{file=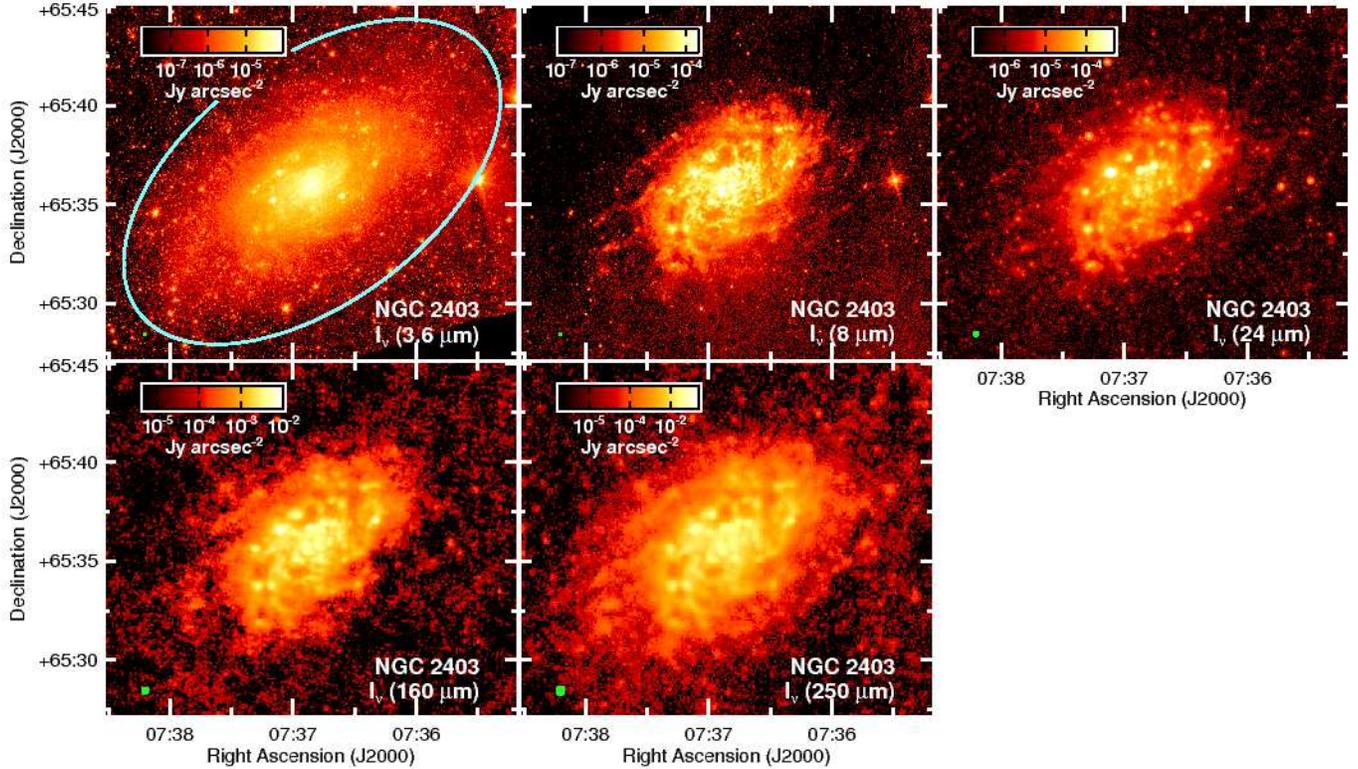}
\caption{The 21 $\times$ 18~arcmin images of NGC 2403 used in the analysis. North is up and east is to the left in each image. The 3.6~$\mu$m maps trace the intermediate-age and older stars. The 8~$\mu$m image mainly shows the PAH 7.7~$\mu$m emission feature but may also contain small amounts of emission from hot dust and stellar sources.  The 24~$\mu$m band traces emission from hot (100~K) dust, and the 160 and 250~$\mu$m trace emission from colder (15-30~K) dust. The FWHM for each image is shown as a green circle in the lower left corner of each panel, and the light blue ellipse in the 3.6~$\mu$m image outlines the optical disk of the galaxy.}
\label{f_ngc2403_maps}
\end{center}
\end{figure*}

\begin{figure*}
\begin{center}
\epsfig{file=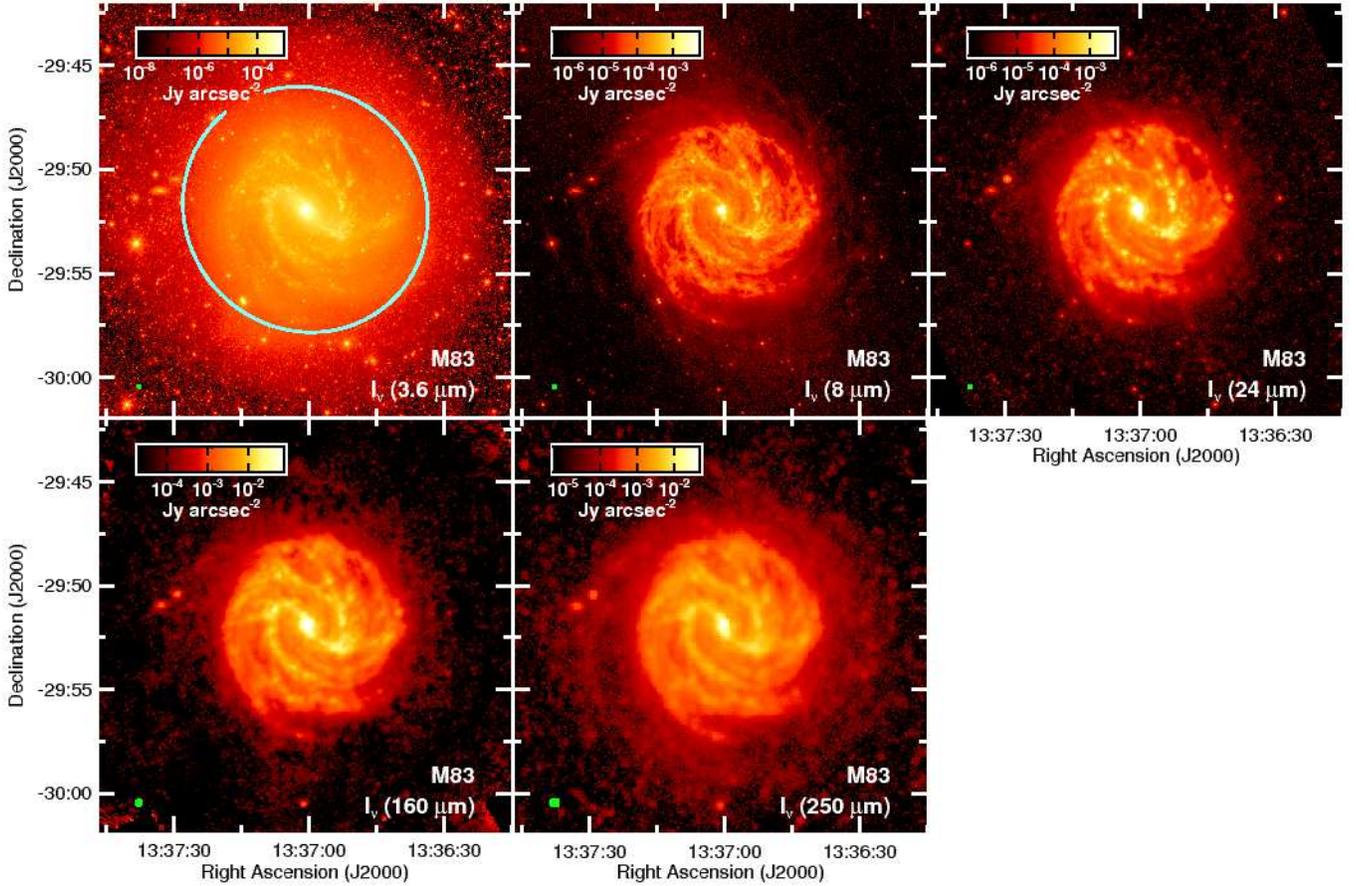}
\caption{The 20 $\times$ 20~arcmin images of M83 used in the analysis.  See Figure \ref{f_ngc2403_maps} for additional information on the image format.}
\label{f_m83_maps}
\end{center}
\end{figure*}

\begin{figure}
\epsfig{file=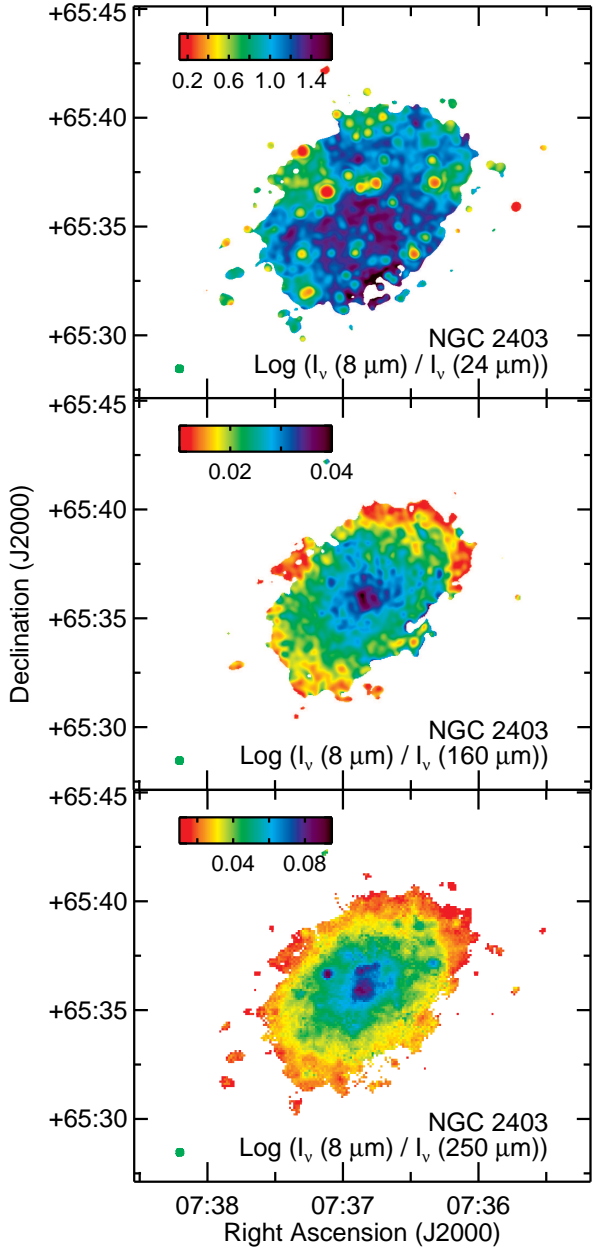}
\caption{The 8/24, 8/160 and 8/250~$\mu$m surface brightness ratio maps for NGC 2403.  These images are based on data where the PSFs are matched to the PSF of the 250~$\mu$m data. The FWHM for the 250~$\mu$m PSF is shown by the green circle in the lower left corner of each panel.  The colour scales in the images have been adjusted to show the structure in the surface brightness ratios; some of the red or purple pixels may be outside the range of values shown in the colour bars.  Data not detected at the $5\sigma$ level in either band is left blank.  The 8/24~$\mu$m ratio is low in star forming regions where the 24~$\mu$m emission is brightest.  The 8/160 and 8/250~$\mu$m images are similar in that the ratios generally decrease with radius.  However, the 8/160~$\mu$m ratio shows more structure, while the 8/250~$\mu$m map is generally smoother.}
\label{f_ngc2403_ratio}
\end{figure}

\begin{figure}
\epsfig{file=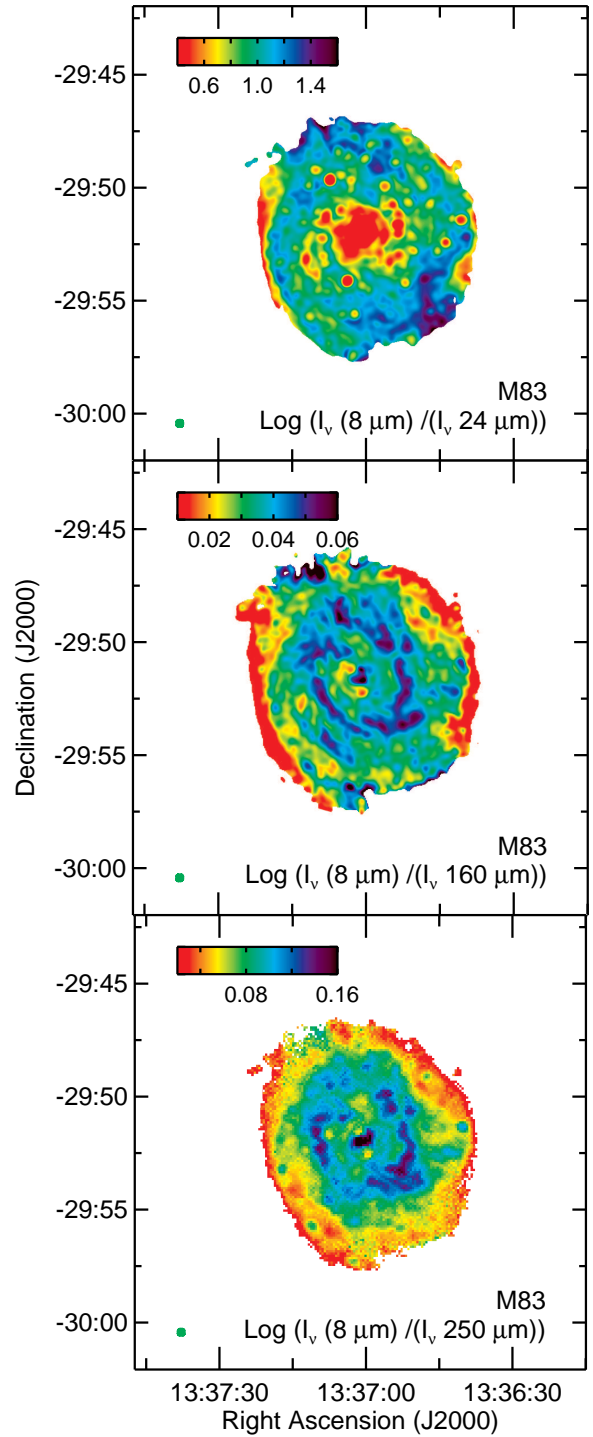}
\caption{The 8/24, 8/160 and 8/250~$\mu$m surface brightness ratio maps for M83.  These images are based on data where the PSFs are matched to the PSF of the 250~$\mu$m data.   The maps are formatted in the same way as the maps in Figure~\ref{f_ngc2403_ratio}.  We see spiral arm structure in all images.  The arm structure in the 8/24~$\mu$m image is traced by a series of red point-like sources where the ratio decreases in star forming regions.  However, the filamentary spiral structures in the 8/160 and 8/250~$\mu$m maps are locations offset from the 160 and 250~$\mu$m emission in Figure~\ref{f_m83_maps} where PAH emission is enhanced relative to cold dust emission.}
\label{f_m83_ratio}
\end{figure}

\begin{figure}
\epsfig{file=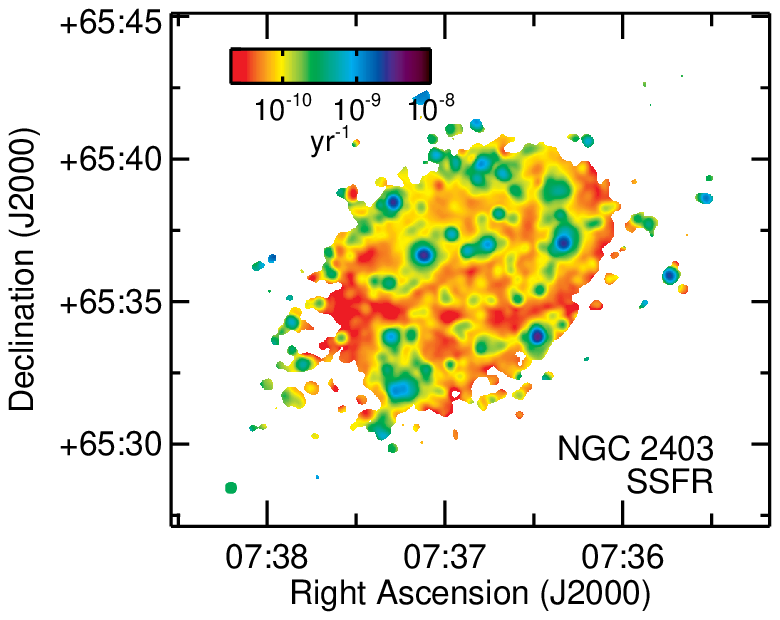}
\epsfig{file=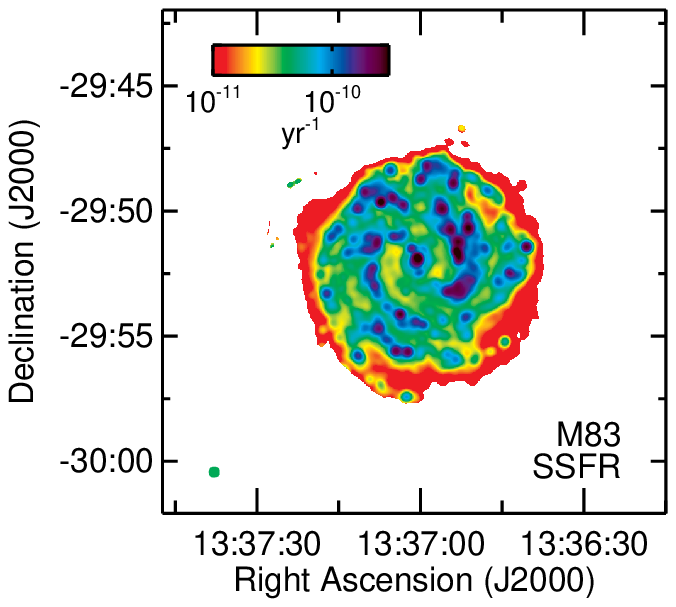}
\caption{ Maps showing the specific star formation rate (SSFR) for NGC~2403 and M83. The data are formatted the same as the corresponding ratio maps for each galaxy in Figures \ref{f_ngc2403_ratio} and \ref{f_m83_ratio}}
\label{f_ssfr}
\end{figure}

Figures \ref{f_ngc2403_maps} and \ref{f_m83_maps} show the 3.6, 8, 24, 160 and 250~$\mu$m images used in the analysis (before the application of the convolution or rebinning steps described in Section~\ref{s_data_prep}).  Figures \ref{f_ngc2403_ratio} and \ref{f_m83_ratio} show the 8/24, 8/160 and 8/250~$\mu$m surface brightness ratios of the data after the convolution step but before rebinning.  The 8, 24, 160, and 250~$\mu$m images all look very similar, demonstrating that the PAHs, hot dust, and cold dust are found in the same large-scale structures.  However, the ratio maps demonstrate how the PAH emission varies with respect to the dust traced by the other infrared bands. For comparison to these figures, we also show maps of the SSFR in Figure \ref{f_ssfr}.

In both galaxies, we see a decrease in the 8/24~$\mu$m ratio in locations where the 24~$\mu$m emission peaks. If the PAH emission was tracing star formation in the same way as the hot dust emission, we would see little variation across the 8/24~$\mu$m ratio maps. The disparity indicates either that the 24~$\mu$m emission is enhanced in regions with high SSFR, that the PAH emission is inhibited where the hot dust emission peaks in the centres of regions with high SSFR, or that both effects occur within the star forming regions. In NGC 2403, we see the 8/24~$\mu$m ratio is higher in the diffuse regions outside the regions with high SSFR, particularly in the southern half of the galaxy.  In M83, we see the enhancement of PAHs relative to the 24~$\mu$m emission in not only the interarm regions but also between high SSFR regions in the spiral arms.  The 8/24~$\mu$m ratio is also very low in the starburst nucleus of M83, as is also seen by Wu et al. (2014, submitted).

The 8/160 and 8/250~$\mu$m ratio maps for NGC~2403 and M83 present different results for each galaxy.  In NGC~2403, the 8/160 and 8/250~$\mu$m ratios peak near the centre and decrease with radius, although the 8/160~$\mu$m map looks more noisy than the 8/250~$\mu$m map.  Instead of seeing the PAH emission decrease relative to the cold dust emission in individual regions with high SSFR, as was the case in the 8/24~$\mu$m ratio maps, we see the PAH emission enhanced relative to the 160 and 250~$\mu$m emission at the location of the infrared-brightest star forming region in the northeast side of the disc. The 8/160 and 8/250~$\mu$m ratios generally do not change significantly near most other star forming regions.  With the exception of the infrared-brightest star forming region, the radial gradients in the 8/160 and 8/250~$\mu$m ratios look similar to the radial gradients in the 3.6~$\mu$m image seen in Figure~\ref{f_ngc2403_maps}.  In NGC~2403, \citet{2012bendo} found that the 160/250~$\mu$m surface brightness ratios were correlated with H$\alpha$ emission and peaked in locations with strong star formation, while the 250/350~$\mu$m ratios were more strongly correlated with near-infrared emission and generally varied radially in the same way as the older stellar populations.  These results demonstrated that the 160~$\mu$m emission is dominated by dust heated locally in star forming regions but the dust seen at 250~$\mu$m is heated by the diffuse ISRF.  The similarity between the 8/160, 8/250, 250/350, and 3.6~$\mu$m radial gradients suggests that the PAHs are intermixed with the cold large dust grains and that the enhancement of PAH emission relative to the large dust grains depends on the surface brightness of the evolved stellar population.  If this is the case, the 8/160~$\mu$m map may looks noisy compared to the 8/250~$\mu$m map because emission in the 8 and 160~$\mu$m bands is affected by different stellar populations while emission in the 8 and 250~$\mu$m bands is affected by mainly the evolved stellar population.

The dust emission for the different wavebands peak in slightly different places in these profiles of the M83's spiral arms.  The 250~$\mu$m emission, which trace most of the dust mass in the spiral arms, peak on the downstream (or inner) side of the spiral arm.  The profile of the 24~$\mu$m emission tends to appear narrower and peaks $0-7$~arcsec further towards the upstream (or outer) side of the spiral arms (although this is small relative to the 18~arcsec resolution of the data used to create these plots).  This is consistent with the classical description of star formation within spiral arms \citep[e.g.]{1969roberts, 1979elmegreen}.  First the gas flows into the spiral arms, then the gas is shocked by the spiral density waves and collapses into stars, and finally young stars emerge on the upstream side of the spiral arms.

\begin{figure*}
\begin{center}
\epsfig{file=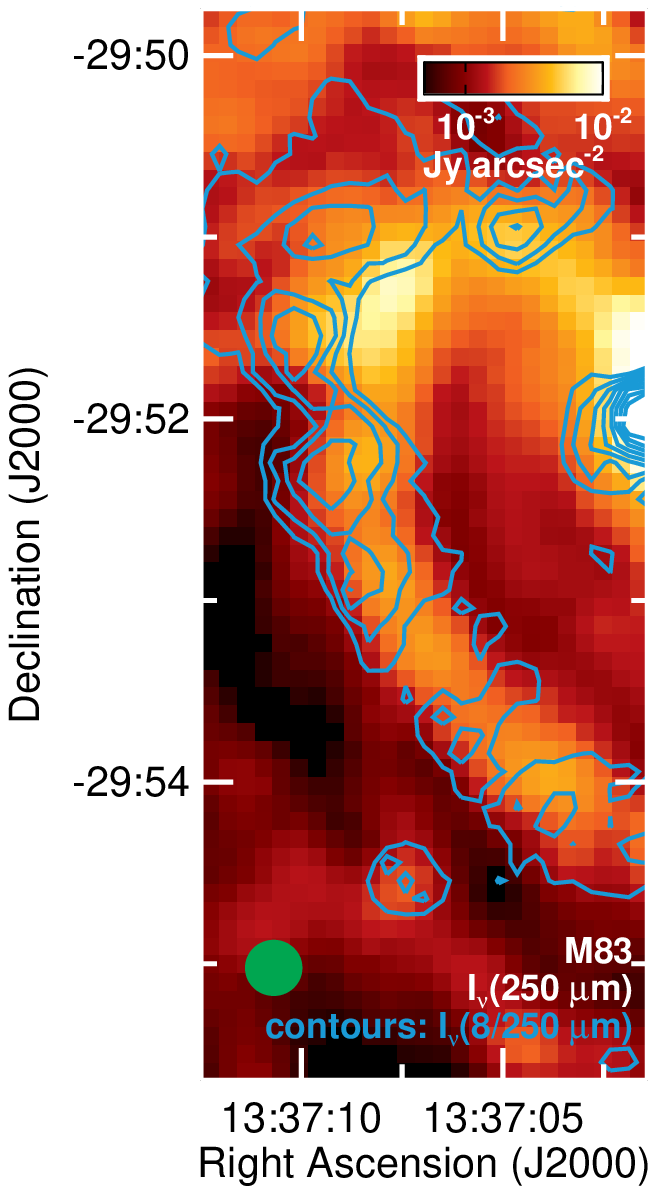,height=7.1cm}
\epsfig{file=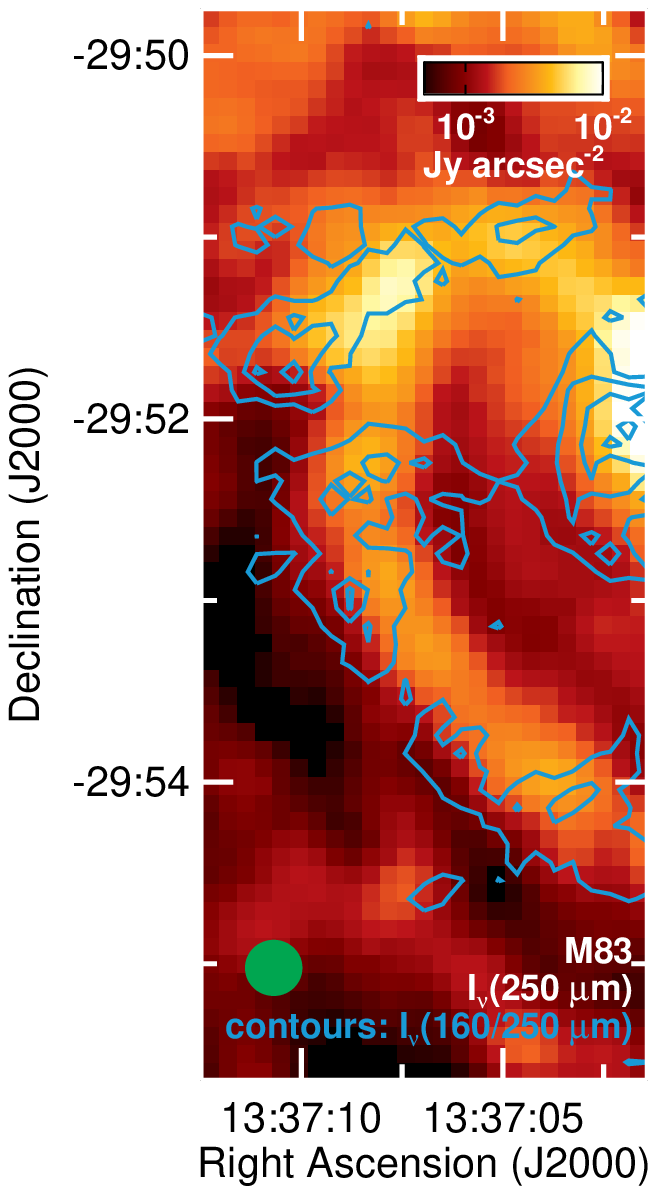,height=7.1cm}
\epsfig{file=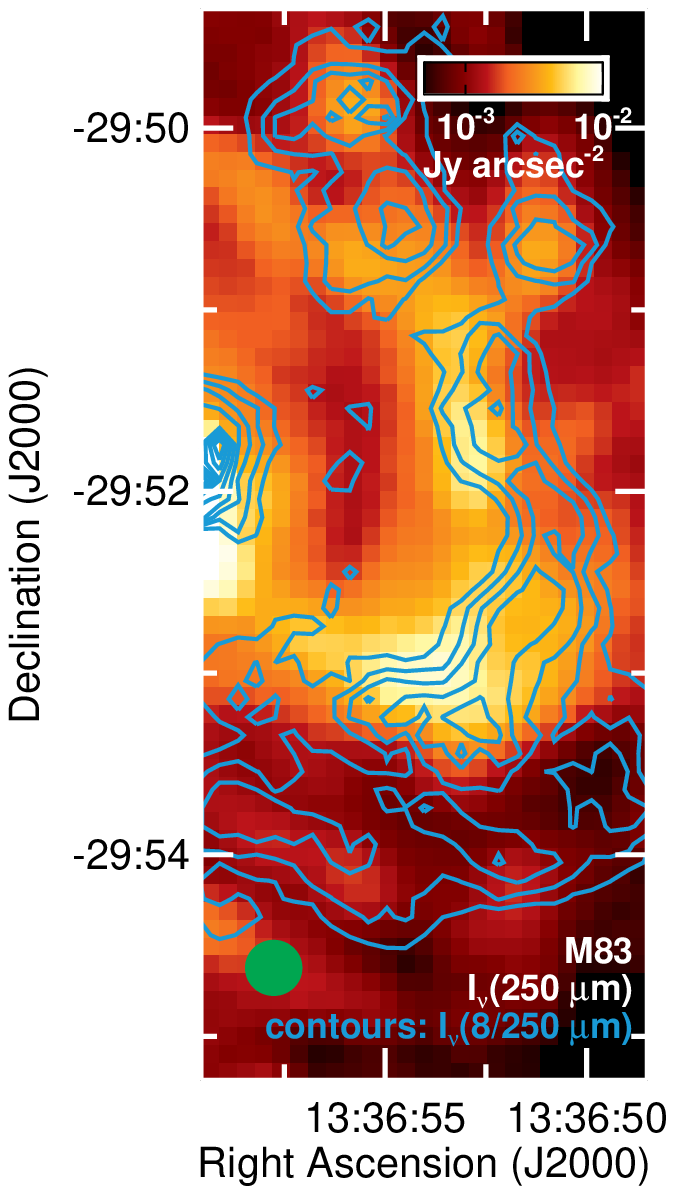,height=7.1cm}
\epsfig{file=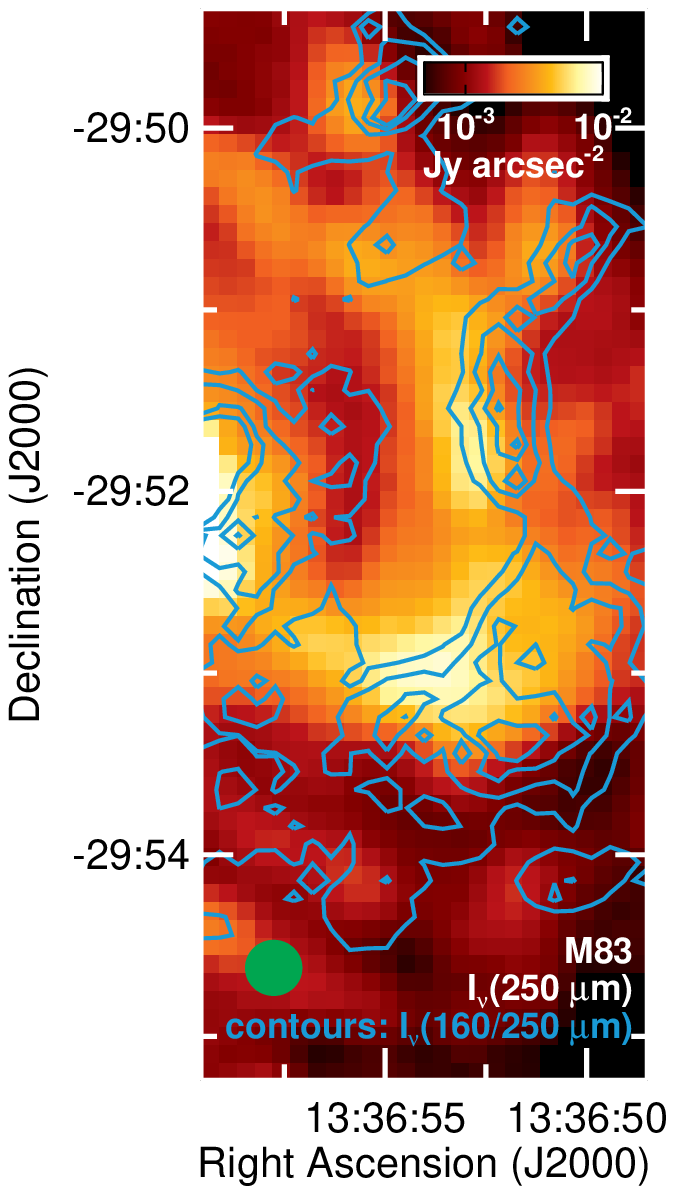,height=7.1cm}
\caption{The 250~$\mu$m images of the two spiral arms in M83 with the 8/250~$\mu$m ratio and 160/250~$\mu$m ratio overlaid as contours.  The contours for the 8/250~$\mu$m ratio start at 0.07 and increase upwards in increments of 0.01.  The contours for the 160/250~$\mu$m ratio start at 2.4 and increase upwards in increments of 0.2.  The images are formatted in the same way as Figure~\ref{f_m83_maps}.  The 8/250~$\mu$m ratios themselves are discussed in Section~\ref{s_analysis_maps}, while both the 8/250 and 160/250~$\mu$m ratios are discussed in Section~\ref{s_m83pah}.}
\label{f_m83_overlay}
\end{center}
\end{figure*}

The 8~$\mu$m emission peaks slightly further downstream from the 250~$\mu$m emission, and the profiles of the 8~$\mu$m emission on the downstream side of the spiral arms is broader than the profiles on the upstream side.  This is particularly pronounced for profiles A, C, and F.  While the 8/24~$\mu$m ratio drops sharply near the star forming regions as expected, the 8/250~$\mu$m ratio peak 10-30~arcsec (or $\sim$200-650 pc) downstream from the dust lane, as is also seen in the countour overlays in Figure~\ref{f_m83_overlay}. This demonstrates that the PAHs emission is enhanced relative to the cold dust on the downstream side of the spiral arms well outside the dust lanes.  The ultraviolet emission also peaks downstream from the 250~$\mu$m emission in many of these profiles and that the profiles of the ultraviolet emission in B and C look broader on the downstream side.  The possible connection of these profiles to the ultraviolet emission and to the 160/250~$\mu$m ratios is discussed further in Section~\ref{s_m83pah}.

The offset enhancements in PAH emission in the arms of M83 could appear because of astrometry problems, but we have checked the astrometry among the images using foreground and background sources outside the optical disc of the galaxy and found no significant offsets greater than $\sim1$~arcsec in the sources between images.  It is also possible that the broader PAH emission could result from issues related to the PSF matching step, but usually these types of artefacts will appear symmetric around bright sources, whereas the enhanced PAH emisson appears asymmetric.  It is more likely that the phenomenon is real and has been difficult to detect before because of limitations in the angular resolution of far-infrared data.

\begin{figure}
\epsfig{file=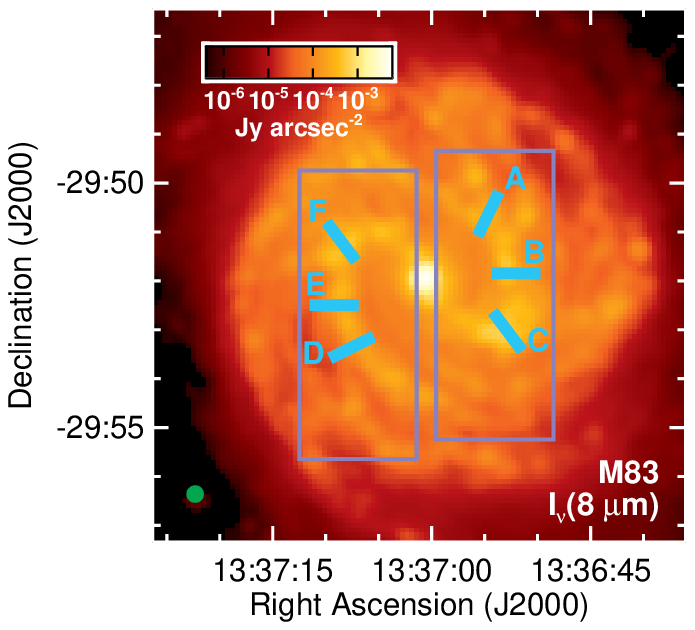}
\caption{The 8~$\mu$m image of M83 (after the PSF has been matched to the PSF of the 250~$\mu$m data) showing locations where we produced additional plots to illustrate the offset between the 8/250~$\mu$m ratio and the dust emission from the spiral arms.  The blue boxes show the locations in Figure~\ref{f_m83_overlay} where we overlay contours of the 8/250~$\mu$m and 160/250~$\mu$m ratios on the 250~$\mu$m data.  The cyan lines show the locations of the surface brightness profiles plotted in Figure~\ref{f_m83_line}.  The image is formatted in the same way as Figure~\ref{f_m83_maps}.}
\label{f_m83_line_map}
\end{figure}

\begin{figure*}
\begin{center}
\epsfig{file=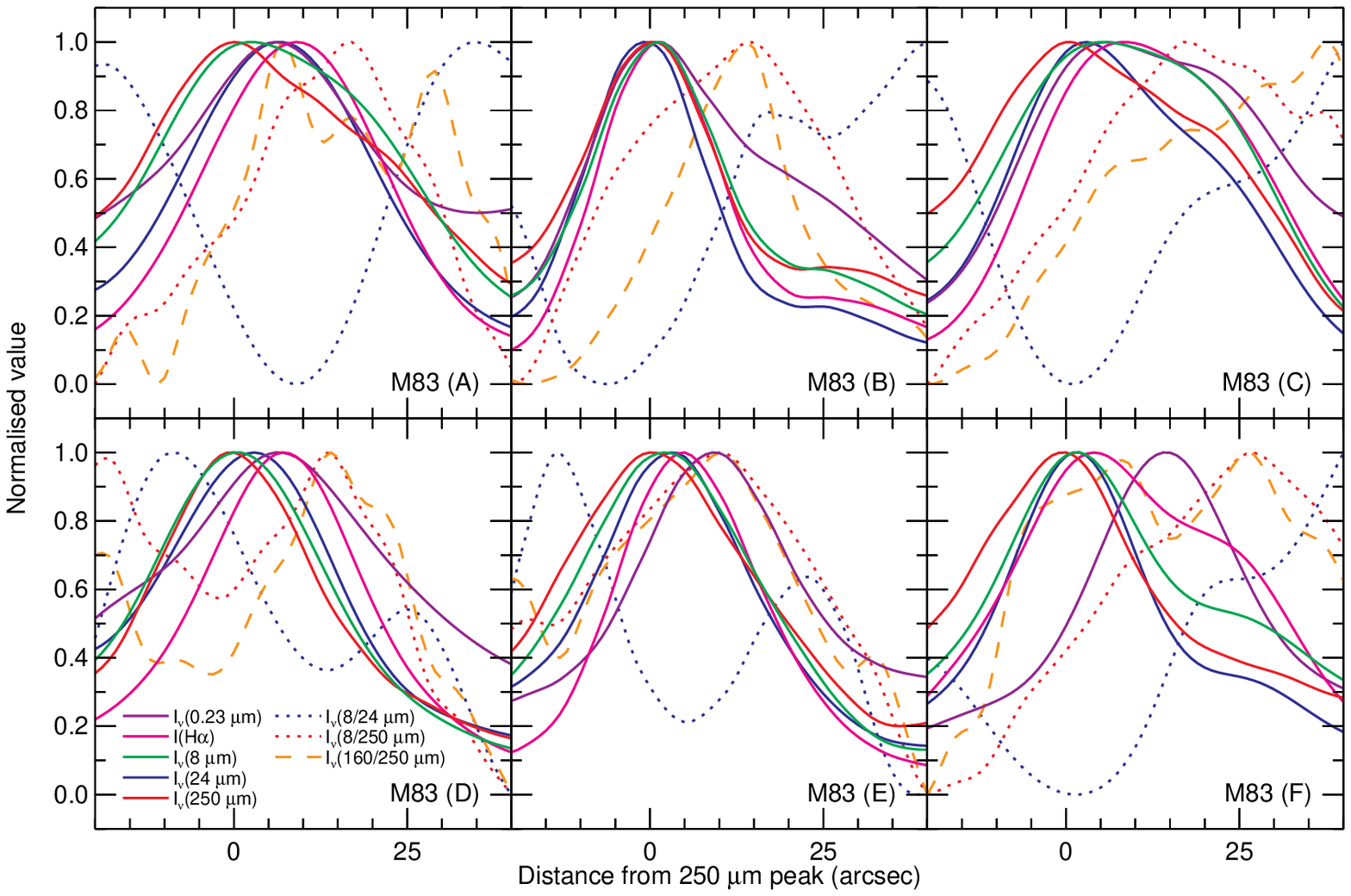}
\caption{Profiles of the H$\alpha$, 0.23, 8, 24, 160, and 250~$\mu$m surface brightnesses and 8/24, {8/250, and 160/250}~$\mu$m surface brightness ratios measured in 18~arcsec wide regions along the locations shown in Figure~\ref{f_m83_line_map}.  The x-axis shows the distance from the peak of the 250~$\mu$m emission; negative numbers are for the upstream side of the arms, while positive materials are for the downstream side.  The profiles were measured in 18~arcsec wide regions in images where the PSF had been matched to the PSF of the 250~$\mu$m data, which has a FWHM of 18~arcsec (or $\sim400$~pc), but the data are supersampled at 1~arcsec resolutions to produce smooth curves.  All surface brightnesses are normalised so that the peak values are 1, and all ratios are normalised so that they range between 0 and 1.  The uncertainties in the normalised surface brightnesses are $\ltsim 1$\%, and the uncertainties in the normalised ratios are $\ltsim 5$\%.  The H$\alpha$ data shown here are corrected for foreground dust attenuation but not corrected for dust extinction within M83.}
\label{f_m83_line}
\end{center}
\end{figure*}

\subsection{Analysis of binned data}
\label{s_binned_analysis}

\begin{figure*}
\begin{center}
\epsfig{file=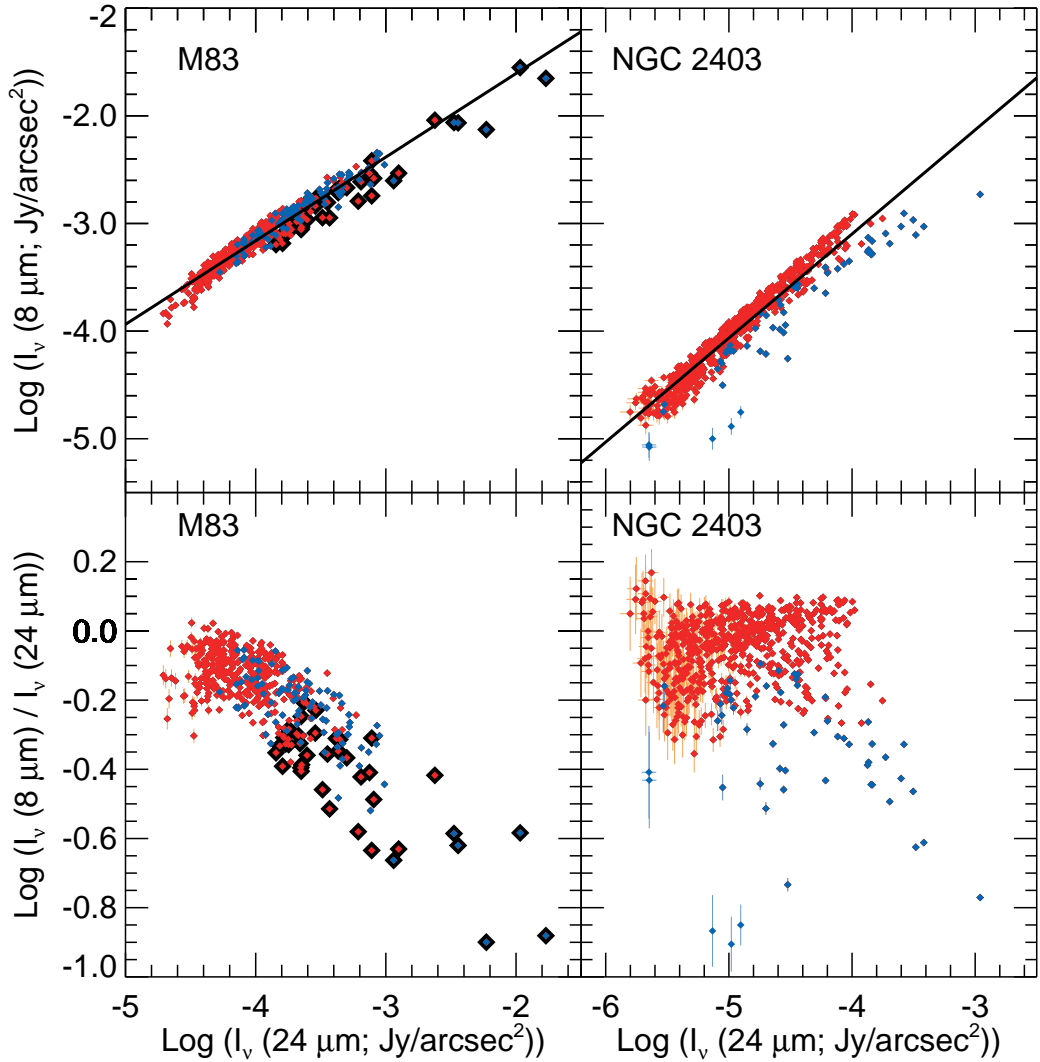}
\caption{The 8~$\mu$m surface brightness and the 8/24~$\mu$m ratios as a function of 24~$\mu$m emission for the 18~arcsec binned data for both galaxies.  Only data detected at the $5\sigma$ level are displayed.  The best fitting linear functions between the surface brightnesses (weighted by the errors in both quantities) are shown as black lines in the top panels. The blue points are locations with high SSFR, and the red points and error bars are locations which are predominantly heated by the diffuse ISRF; see Section~\ref{s_data_prep} for more details. In M83 we highlight locations within a 1.5~kpc radius of the centre in black. }
\label{f_8-24}
\end{center}
\end{figure*}

\begin{figure*}
\begin{center}
\epsfig{file=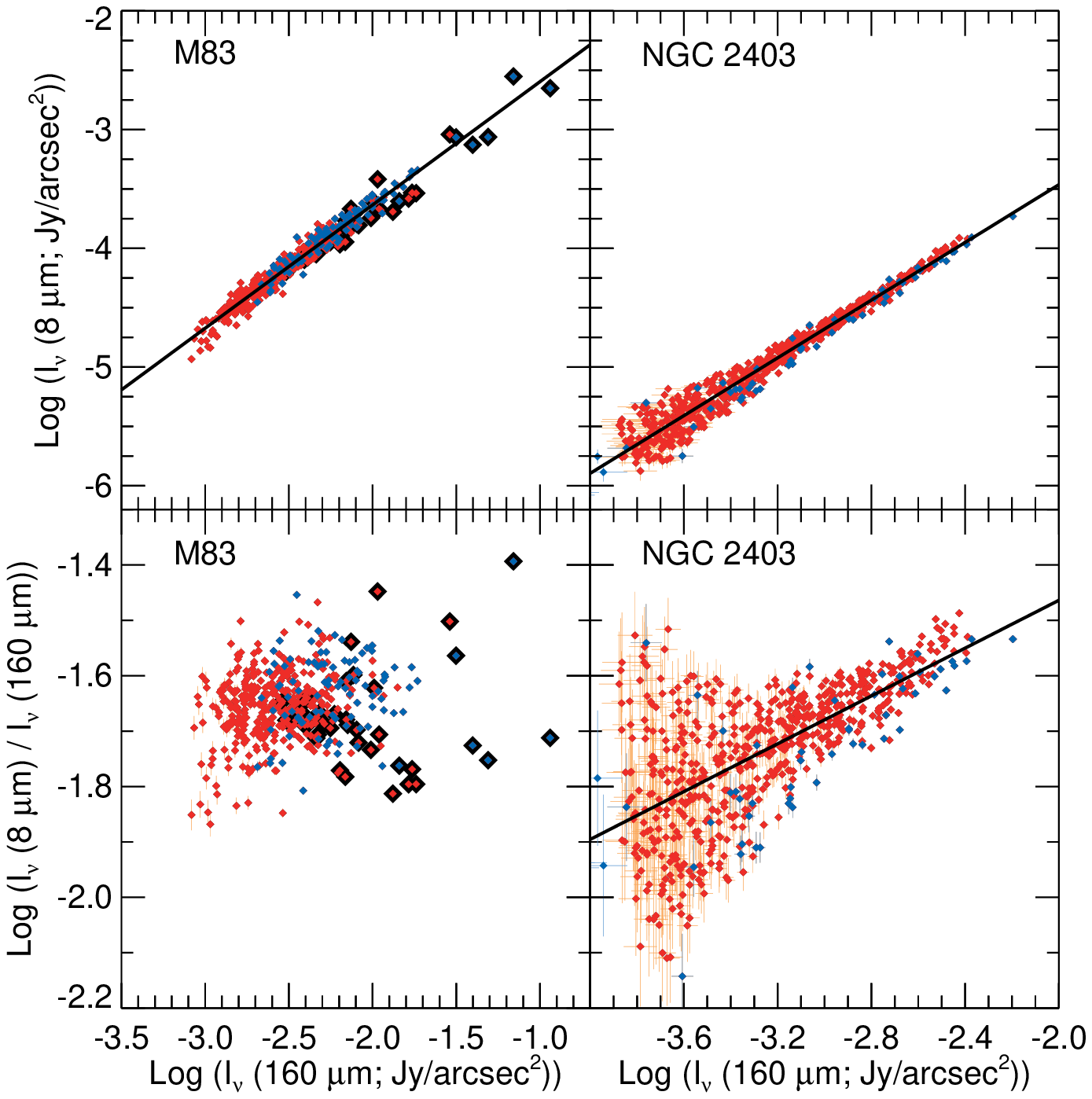}
\caption{The 8~$\mu$m surface brightness and the 8/160~$\mu$m ratios as a function of 160~$\mu$m emission for the 18~arcsec binned data for both galaxies.  The data are formatted in the same way as in Figure~\ref{f_8-24}.}
\label{f_8-160}
\end{center}
\end{figure*}

\begin{figure*}
\begin{center}
\epsfig{file=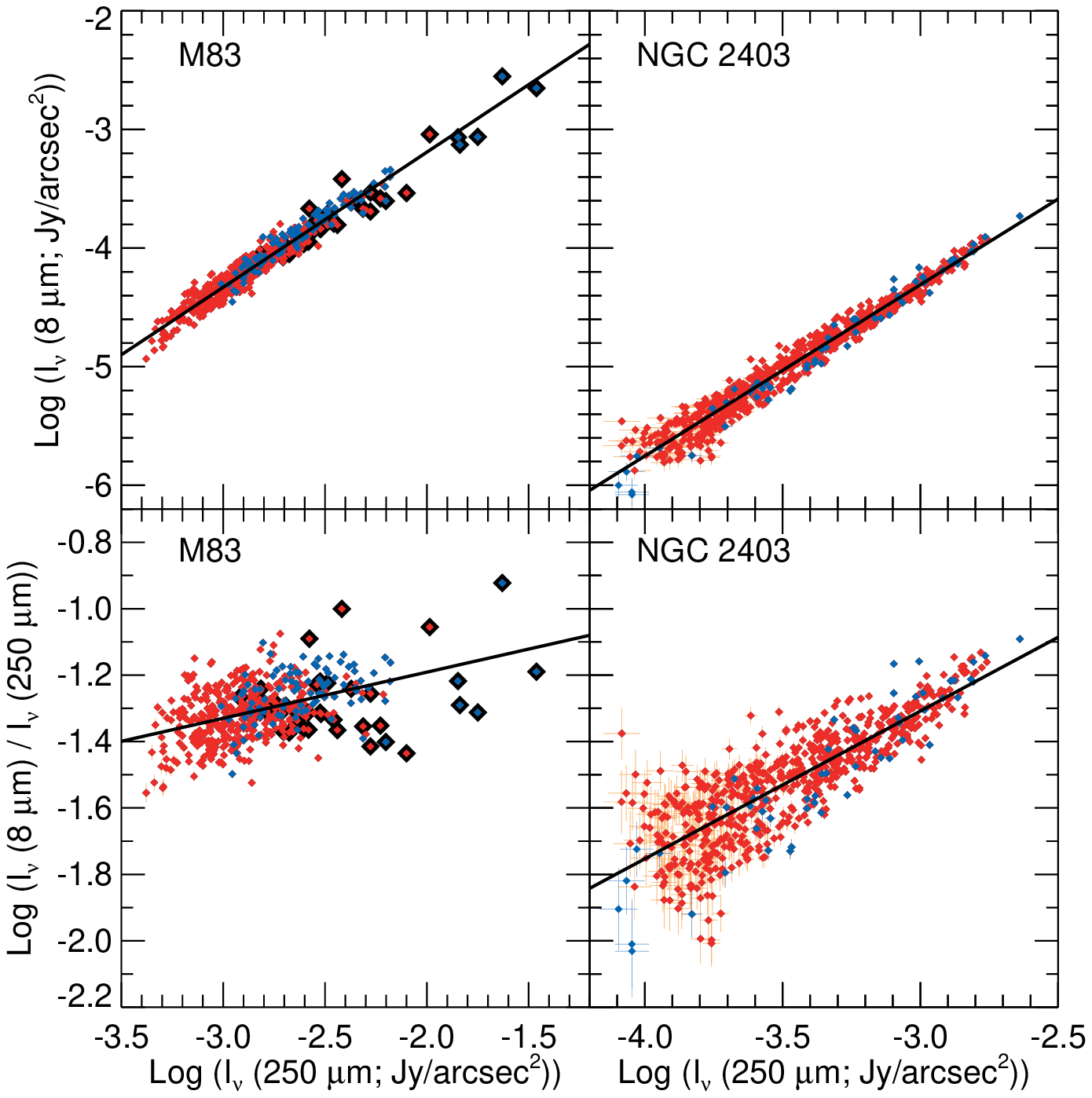}
\caption{The 8~$\mu$m surface brightness and the 8/250~$\mu$m ratios as a function of 250~$\mu$m emission for the 18~arcsec binned data for both galaxies.  The data are formatted in the same way as in Figure~\ref{f_8-24}.}
\label{f_8-250}
\end{center}
\end{figure*}

In Figures \ref{f_8-24}-\ref{f_8-250}, we plot the relations between the PAH emission at 8~$\mu$m and either the hot dust emission at 24~$\mu$m or the cold dust emission at 160 and 250~$\mu$m.  Pearson correlation coefficients for these relations are given in Table \ref{t_correlation}.  To first order, the 8~$\mu$m surface brightness is well correlated with the 24, 160 and 250~$\mu$m surface brightnesses.  However, the 8/24, 8/160 and 8/250~$\mu$m ratios reveal the presence of both scatter in the relations between the 8~$\mu$m emission and emission in other bands as well as systematic variations in these relations.

The variations in the 8/24~$\mu$m ratio in Figure~\ref{f_8-24} show that the 8/24~$\mu$m ratio decreases in areas where the 24~$\mu$m emission is strongest in both galaxies.  However, the relations between the 8/24~$\mu$m ratio and the 24~$\mu$m emission differ somewhat between the two galaxies.  In NGC~2403, we found that we could see different trends in the data when we separated the 18~arcsec binned data into two subsets where the SSFR was either $\geq1\times10^{-10}$~yr$^{-1}$ or $<1\times10^{-10}$~yr$^{-1}$.  The data with low SSFR follow a relation in which $\log (I_\nu(8\mu\mbox{m})/I_\nu(24\mu\mbox{m}))$ increases slightly from $\sim 0.0$ to $\sim 0.2$ as $\log (I_\nu(24\mu\mbox{m}))$ increases from -6 to -4.  The emission from these regions may orginate mainly from locations in the diffuse ISM with relatively soft radiation fields where the PAH emission is well-correlated with emission from hot, diffuse dust heated by the diffuse ISRF.  As both the PAHs and the hot, diffuse dust are stochastically heated, the ratio of PAH to hot dust emission is expected to be roughly constant.  The slight decrease in the 8/24~$\mu$m ratio as the diffuse 24~$\mu$m surface brightness decreases is potentially a result of an increase in the hardness of the radiation field as the 24~$\mu$m surface brightness decreases (possibly as a result of changes in metallicity with radius as found by multiple authors \citep[e.g. ][]{1994zaritsky,2010moustakas}), which could lead to either PAH emission being suppressed or 24~$\mu$m emission being enhanced.  Data points tracing locations with high SSFR fall below the relation between the 8/24~$\mu$m ratio and 24~$\mu$m surface brightness.  These locations would be expected to have harder radiation fields that may enhance the 24~$\mu$m emission or suppress the PAH emission.  While we are able to empirically separate data into regions with high and low 8/24~$\mu$m ratios using a SSFR cutoff value of $1\times10^{-10}$~yr$^{-1}$, additional work in modelling the stellar populations, the PAH excitation, and the dust heating is needed to understand the details of how the SSFR of the stellar populations affects the variations in the 8/24~$\mu$m ratio within NGC~2403.

In M83, the relationship between the 8/24~$\mu$m ratio and 24~$\mu$m emission is close to linear at $\log (I_\nu(24\mu\mbox{m}))>-4$, but it flattens at $\log (I_\nu(24\mu\mbox{m})) < -4$. Some of the lowest 8/24~$\mu$m ratios corresponds to regions with high SSFR in the nucleus and spiral arms where, again, the 24~$\mu$m emission may be strongly enhanced or the PAH emission is suppressed.  The relation between 24~$\mu$m emission and the 8/24~$\mu$m ratio at $\log (I_\nu(24\mu\mbox{m}))>-4$ in M83 is similar to the relation seen in NGC~2403.  Unlike NGC~2403, however, we found that we could not readily separate data in the plots of $\log (I_\nu(8\mu\mbox{m})/I_\nu(24\mu\mbox{m}))$ versus $\log (I_\nu(24\mu\mbox{m}))$ for M83 simply by selecting data by SSFR, as some regions with low SSFR have low 8/24~$\mu$m ratios.  These regions are mostly locations within radii of 1.5~kpc.  The exact reason why we see this is unclear, although it is possible that hard ultraviolet photons from the starburst nucleus leak into the diffuse ISM in this region and destroy the PAHs in the diffuse ISM.

Figure \ref{f_8-160} shows good correlations between the 8~$\mu$m and 160~$\mu$m surface brightnesses.  The plot of the 8/160~$\mu$m vs 160~$\mu$m for M83 shows that the 8/160~$\mu$m ratio is close to constant over a range of infrared surface brightnesses that vary by a factor of 100, indicating that the relation between 8 and 160~$\mu$m emission is very close to a one-to-one relationship.  Some scatter is seen in the 8/160~$\mu$m ratio at high 160~$\mu$m surface brightnesses.  Some of these data points are for locations around the infrared-bright centre of M83 where the outer regions of the PSF were not matched perfectly in the convolution step, while other data points sample regions along the spiral arms where the enhancement in the 8/160~$\mu$m ratio is offset from the 160~$\mu$m surface brightness as discussed in Section~\ref{s_analysis_maps}.  In NGC 2403, the 8/160~$\mu$m ratio increases with 160~$\mu$m surface brightness, and the relation exhibits more scatter, indicating that the relation of 8~$\mu$m emission to 160~$\mu$m emission in NGC~2403 is different from the relation for M83.

The relations between the 8 and 250~$\mu$m emission in Figure \ref{f_8-250} are similar to the relations between the 8 to 160~$\mu$m emission.  For both galaxies, the 8/250~$\mu$m ratio increases with the 250~$\mu$m surface brightness, and the correlation coefficients are relatively strong.  In NGC~2403, the correlation coefficient between the 8/250~$\mu$m ratio and the 250~$\mu$m surface brightness is 0.83, which is much higher than the correlation coefficient of 0.66 for the relation between the 8/160~$\mu$m ratio and the 160~$\mu$m surface brightness.  Given that the square of the Pearson correlation coefficient indicates the fraction of variance in one quantity that depends upon the other quantity, the difference in the correlation coefficients is equivalent to a $\sim25$\% difference in being able to describe the variance in the relations.  This suggests that the PAHs are more strongly associated with the colder dust seen at 250~$\mu$m than the warmer dust seen at 160~$\mu$m.  In M83, the relation between the 8/250~$\mu$m ratio and 250~$\mu$m surface brightness is sloped and also shows significant scatter at high surfaces brightnesses in the same way as the relationship between the 8/160~$\mu$m ratio and the 8~$\mu$m emission.

Because M83 is at a distance $\sim$1.5 further than NGC 2403, the 18~arcsec bins used in this analysis will cover regions with different spatial scales.  In Appendix~\ref{a_binsizecheck}, we examined how the results for the analysis on NGC~2403 would change if we used 27~arcsec bins, which cover approiximately the same spatial scales as the 18~arcsec bins used for the M83 data. We see no noteable difference in the results using the 27~arcsec bins compared to the 18~arcsec bins; most correlation coefficients change by $\leq0.05$.  Hence, adjusting the bin sizes for the two galaxies to similar spatial scales is unimportant.  We will therefore use data measured in the smaller bins as it takes full advantage of the capabilities of the {\it Herschel} data that we are using and as it allows us to illustrate how the relations are still found in smaller structure in NGC 2403.

\begin{table}
\caption{Pearson correlation coefficients for the binned data.}
\label{t_correlation}
\begin{tabular}{p{5.6cm}cc}
\hline
       & 
   	NGC &
	M83 
	\\ 
       & 
   	2403 &
	\\ \hline
$\log (I_\nu(8\mu\mbox{m}))$ vs $\log(I_\nu(24\mu\mbox{m}))$ &
   	0.96	&
	0.97	\\
$\log (I_\nu(8\mu\mbox{m})/I_\nu(24\mu\mbox{m}))$ vs $\log(I_\nu(24\mu\mbox{m}))$ &
	-0.13	&
	-0.74	\\
$\log (I_\nu(8\mu\mbox{m}))$ vs $\log(I_\nu(160\mu\mbox{m}))$ &
	0.98	&
	0.98	\\ 
$\log (I_\nu(8\mu\mbox{m})/I_\nu(160\mu\mbox{m}))$ vs $\log(I_\nu(160\mu\mbox{m}))$ &	
	0.66	&
	0.17	\\
$\log (I_\nu(8\mu\mbox{m}))$ vs $\log(I_\nu(250\mu\mbox{m}))$ &
	0.98	&	
	0.97	\\ 
$\log (I_\nu(8\mu\mbox{m})/I_\nu(250\mu\mbox{m}))$ vs $\log(I_\nu(250\mu\mbox{m}))$ &
	0.83	&
	0.44	\\		
	\hline
\end{tabular}	
\end{table}

\section{Identification of PAH excitation sources}
\label{s_pahexcitation}

\subsection{PAH excitation in NGC 2403}
\label{s_ngc2403pah}

We can conclude from our analysis that the 8~$\mu$m emission that we observe from NGC~2403 does not originate from PAHs excited locally within the centres of star forming regions, as the relationship between 8 and 24~$\mu$m emission shows that PAH 8~$\mu$m emission decreases relative to 24~$\mu$m emission within regions with high SSFR.  This is partly because the 24~$\mu$m emission is very sensitive to dust heating and increases significantly within star forming regions \citep[e.g.][]{2001dale, 2002dale}.  However, the strong ultraviolet radiation from these massive young stars probably photodissociates the PAHs in the centres of these regions including species that produce features other than the 7.7~$\mu$m feature, as has been seen spectroscopically in other galactic and extragalactic star forming regions \citep{2007berne, 2007lebouteiller, 2007povich, 2008gordon}.  PAH emission has been observed in the outer regions of photodissociation regions \citep{2007berne, 2007lebouteiller, 2007povich}; the photons that excite the PAHs in these locations can also heat the very small grains that produce the 24~$\mu$m emission.  In our data, the emission from the inner and outer regions of these regions will be blended.  Integrating over the centres of these regions, the 8/24~$\mu$m ratio will still appear low comapred to diffuse regions outside these regions because the PAH emission is suppressed in parts of the regions while the 24~$\mu$m emission is not, a result also obtained by \citet{2005calzetti}.  The dust emitting at 160~$\mu$m is also heated by light from star forming regions \citep{2012bendo}.  While the 8~$\mu$m emission is better correlated with the 160~$\mu$m band than with the 24~$\mu$m band, the 8/160~$\mu$m ratio still shows significant scatter as a function of 160~$\mu$m surface brightness, possibly because the PAH emission is still inhibited in the regions from which the 160~$\mu$m emission is originating.

However, we see a strong correlation between the 8 and 250~$\mu$m surface brightnesses, and the relation between the 8/250~$\mu$m ratio and the 250~$\mu$m surface brightness shows that the residuals in the relation between the 8 and 250~$\mu$m are very small, especially compared to the equivalent residuals for the relations between the 8~$\mu$m emission and emission in either the 24~$\mu$m or 160~$\mu$m bands. This is particularly evident when comparing the correlation coefficients for the 8/24~$\mu$m ratio versus 24~$\mu$m emission, the 8/160~$\mu$m versus 160~$\mu$m emission, and the 8/250~$\mu$m ratio versus 250~$\mu$m emission in Table~\ref{t_correlation}.  This indicates that the PAH emission is much more strongly tied to the dust emitting in the 250~$\mu$m band.  \citet{2012bendo} demonstrated that the dust emitting at $\geq 250$~$\mu$m in NGC~2403 was heated mainly by the diffuse ISRF from the total stellar population.  This implies that the PAHs in NGC~2403 are also mainly excited by the diffuse ISRF.  Moreover, the map of the 8/250~$\mu$m ratio in Figure~\ref{f_ngc2403_ratio} looks very similar to both the 3.6~$\mu$m map in Figure~\ref{f_ngc2403_maps} that traces the light from the total stellar population and the 250/350~$\mu$m ratio map from \citet{2012bendo} that shows the variations in the colour temperatures of the large dust grains heated by the ISRF from these stars.

\begin{figure}
\begin{center}
\epsfig{file=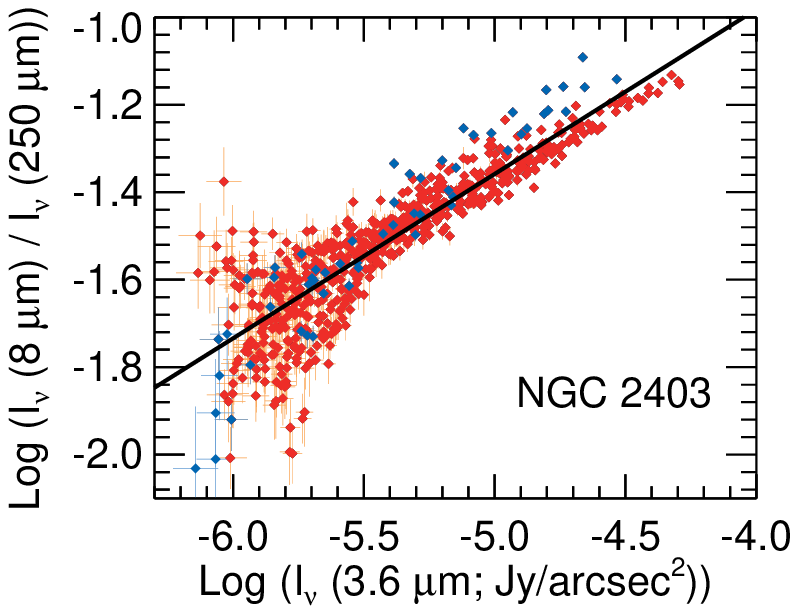}
\epsfig{file=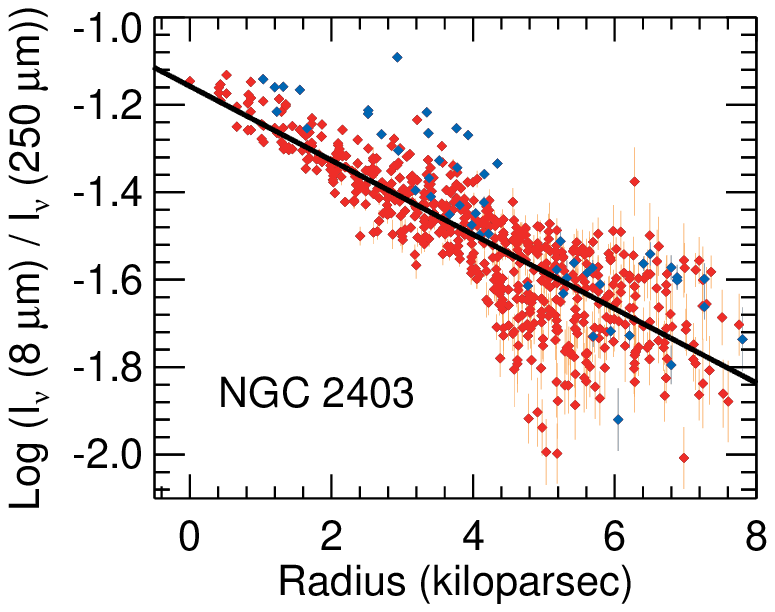}
\caption{The 8/250$~\mu$m surface brightness ratio plotted as a function of the 3.6~$\mu$m surface brightness and galactocentric radius for the 18~arcsec binned data for NGC~2403. The data are formatted in the same way as in Figure~\ref{f_8-24}.  The radii are based on using an inclination of $62.9\deg$ from \citet{2008deblok}.}
\label{f_2403-8-250}
\end{center}
\end{figure}

To examine this relation further, we plot the 8/250~$\mu$m ratio versus the 3.6~$\mu$m surface brightness for NGC 2403 in Figure \ref{f_2403-8-250}. We find a strong correlation between 8/250~$\mu$m ratio and the 3.6~$\mu$m surface brightness; the Pearson correlation coefficient for the relation between these data in logarithmic space is 0.89.  This shows that the enhancement of PAH emission relative to cold dust emission scales with the stellar surface brightness, which implies that the PAHs are primarily mixed in with the large dust grains in the diffuse ISM and that the PAHs are predominantly heated by the diffuse ISRF from the total stellar population.

\begin{figure}
\begin{center}
\epsfig{file=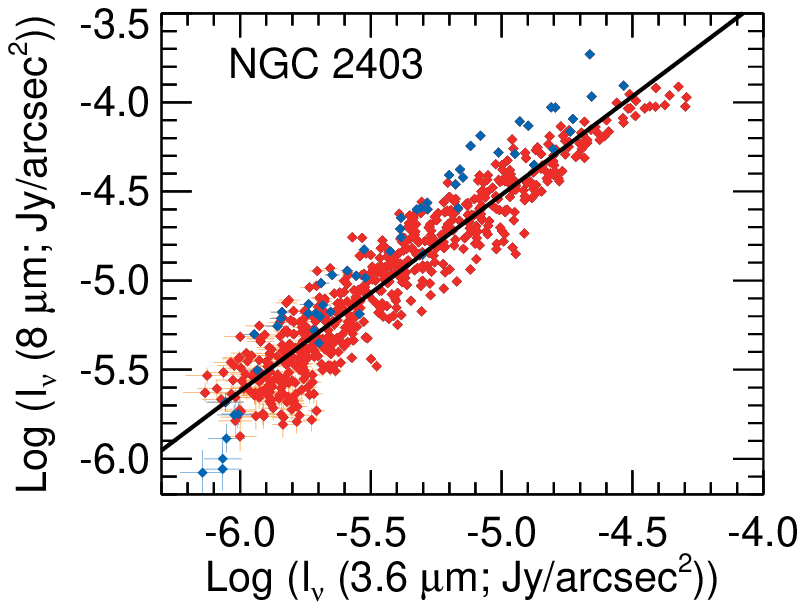}
\epsfig{file=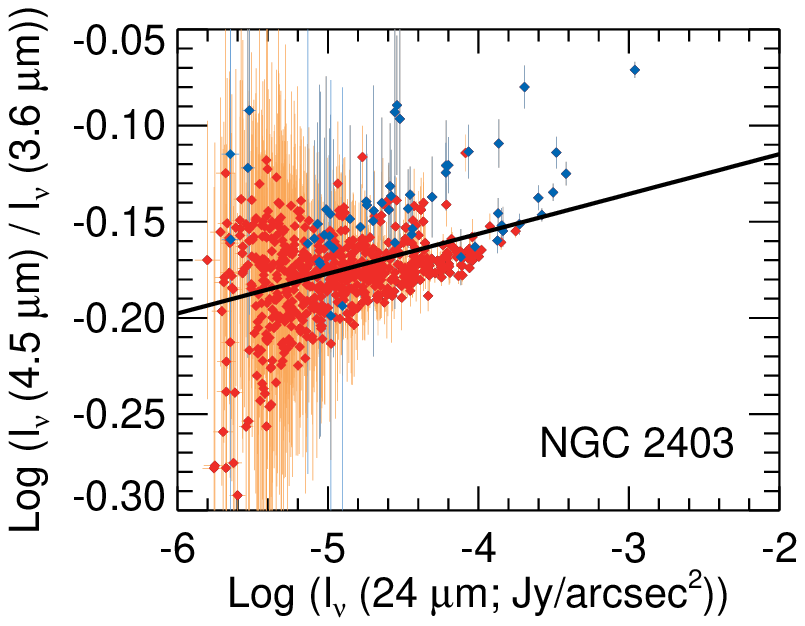}
\caption{To examine whether the relation between the 3.6~$\mu$m emission and the 8/250~$\mu$m ratio is related to hot dust or PAH emission in the 3.6~$\mu$m band, we first show the relation between 3.6 and 8~$\mu$m emission.  The relation in the top panel could be the result of either the 3.6 and 8~$\mu$m bands tracing emission from similar sources or the 3.6~$\mu$m band tracing starlight exciting the PAHs seen in the 8~$\mu$m band.  Next, we plot the 4.5/3.6$~\mu$m ratio as a function of the 24$~\mu$m emission in the bottom panel to examine whether the slope of the SED at 3.6-4.5~$\mu$m is influenced by hot dust emission.  The absence of such a relation as well as the relative invariance of the 4.5/3.6~$\mu$m ratio implies that the 3.6 and 4.5~$\mu$m bands are relatively unaffected by non-stellar emission.  The data are formatted in the same way as in Figure~\ref{f_8-24}.}
\label{f_2403_stars}
\end{center}
\end{figure}

It is also possible that, because the correlation coefficient for the relation between the logarithm of the 3.6~$\mu$m surface brightness and radius is -0.95, the 8/250~$\mu$m ratio actually depends on radius rather than 3.6~$\mu$m surface brightness.  Any radial dependence would be expected to be related to metallicity, which decreases with radius, and this could influence the PAHs.  However, the results from \citet{2008engelbracht} and \citet{2008gordon} indicated that the apparent dependence of PAH emission on metallicity is really the result of changes in the radiation field illuminating the ISM in lower metallicity environments.  The radiation field would be harder in low metallicity systems first because of increases in the stellar temperatures of O and B stars \citep{2004massey, 2005massey, 2007trundle} and second because of decreases of extinction related to a decrease in the gas to dust ratio. This in turn leads to harder interstellar radiation fields in low-metallicity environments that potentially destroy PAHs.  To examine this, we plot the 8/250~$\mu$m ratio as a function of radius in the bottom panel of Figure~\ref{f_2403-8-250}.  As expected, the 8/250~$\mu$m ratio decreases as radius increases.  The relation between the logarithm of the 8/250~$\mu$m ratio and radius has a Pearson correlation coefficient of -0.82, which has an absolute value that is similar to the value of 0.89 for the relation between the lorarithms of the 8/250~$\mu$m ratio and the 3.6~$\mu$m emission.  However, at the resolution of these data, we do see non-axisymmetric substructures in the 8/250~$\mu$m image in Figure~\ref{f_ngc2403_ratio} that correspond to similar substructure in the 3.6~$\mu$m images, but these structures are mainly visible at radii of $<10$ kpc.  When we look at data within this inner 10~kpc region, we get a correlation coefficient of -0.77 for the relation with radius and 0.91 for the relation with 3.6~$\mu$m surface brightness.  This implies that the total stellar surface brightness is more influentual on PAH excitation than any effects related to radius.

Although it is unlikely, the correlation between the 8/250~$\mu$m ratio and 3.6~$\mu$m emission could be the result of a correlation between emission in the 3.6 and 8~$\mu$m bands themselves.  Both bands may contain thermal continuum dust emission (although the thermal continuum emission should have been removed from the 8~$\mu$m data when we applied Equation~\ref{e_pahdustsub}), and the 3.6~$\mu$m band may also contain emission from the 3.3~$\mu$m PAH emission feature, although previous work by \citet{2003lu} indicated that emission at $<5$~$\mu$m from nearby galaxies is dominated by stellar emission.  To investigate this further, we plot the relation between the 8~$\mu$m PAH emission and 3.6~$\mu$m stellar surface brightness in Figure~\ref{f_2403_stars}.  The data are well correlated; the correlation coefficient for the relation in logarithm space is 0.94.  While this could indicate that the same emission sources are seen at 3.6 and 8~$\mu$m, it is also possible that the 8~$\mu$m emission is correlated with 3.6~$\mu$m emission because the stars seen at 3.6~$\mu$m excite the PAHs seen at 8~$\mu$m.  Hence, the correlation between 3.6 and 8~$\mu$m does not necessarily prove anything about the relation between the emission in these bands.  \citet{2010mentuch} illustrated that it was possible to identify the influence of non-stellar emission at near-infrared wavenegths by examining the 4.5/3.6~$\mu$m surface brightness ratio.  This ratio would be relatively invariant for stellar emission because it traces emission from the Rayleigh-Jeans side of the stellar SED, but if hot dust emission influences the bands, the ratio should increase.  We plot the 4.5/3.6~$\mu$m ratio versus 24~$\mu$m emission in the lower panel of Figure \ref{f_2403_stars}. This relationship is almost flat.  Most of the data points have log($I_\nu$(4.5~$\mu$m)/$I_\nu$(3.6~$\mu$m)) values that lie within a range of -0.14 to -0.23, which would be consistent with what was observed for evolved stellar populations by \citet{2010mentuch}.  The absence of significant variations in the 4.5/3.6~$\mu$m ratio with 24~$\mu$m implies that the 3.6 and 4.5~$\mu$m bands are largely uninfluenced by hot dust emission.  We do see a few data points with values of log($I_\nu$(4.5~$\mu$m)/$I_\nu$(3.6~$\mu$m))$>-0.14$ where the 3.6 and 4.5~$\mu$m may be more strongly influenced by non-stellar emission, but these data only weakly influence our results.  If we exclude these data, the correlation coefficient for the relation between the 3.6~$\mu$m data and the 8/250~$\mu$m data changes by $<0.01$.  This shows that the 3.6~$\mu$m emission in NGC~2403 is largely dominated by the stellar population and is relatively unaffected by hot dust or PAH emission. Therefore, the most likely explanation for the correlation between the 3.6 and 8~$\mu$m emission as well as the correlation between the 3.6~$\mu$m emission and the 8/250~$\mu$m ratio is that the PAHs are excited by the stellar population seen at 3.6~$\mu$m.

\subsection{PAH excitation in M83}
\label{s_m83pah}

\begin{figure}
\begin{center}
\epsfig{file=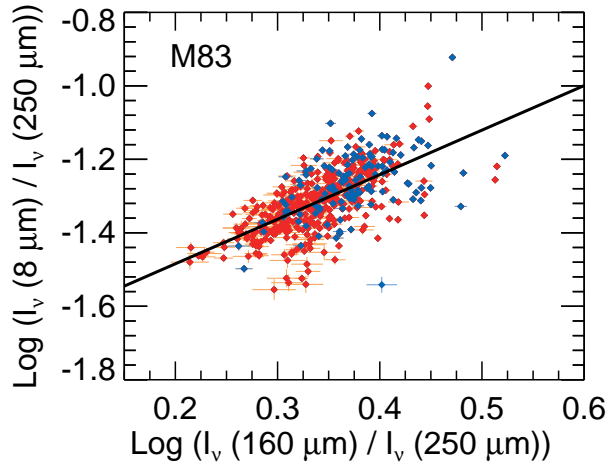}
\caption{The 8/250$~\mu$m surface brightness ratio plotted as a function of the 160/250~$\mu$m surface brightness ratio for the 18~arcsec binned data. The data are formatted in the same way as in Figure~\ref{f_8-24}.}
\label{f_8-250-160-250}
\end{center}
\end{figure}

The results from the 8~$\mu$m to 24~$\mu$m relationship in M83 are similar to NGC 2403.  We see the 8~$\mu$m PAH emission is low in regions where the 24~$\mu$m emission peaks.  Again, PAHs are probably being destroyed locally in regions with high SSFR.  However, we find that 8~$\mu$m emission is more strongly related to the 160 and 250~$\mu$m emission.  We also see offsets between the 8/250~$\mu$m ratios and the dust mass (as traced by the 250~$\mu$m band).  \citet{2012bendo} and \citet{2014bendo} also found that the 160/250~$\mu$m colours appeared offset relative to the star forming regions in the spiral arms, and \citet{2012foyle} found a related offset in the dust colour temperatures.  This implies that the enhancement in PAH emission relative to cold dust emission is related to the enhancement of the temperature of the dust seen at 160~$\mu$m.  To examine this relationship further, we map the 250~$\mu$m emission from the spiral arms overlaid with contours showing the 160/250~$\mu$m ratio in Figure~\ref{f_m83_overlay}, and we show profiles of the 160/250~$\mu$m ratio across the spiral arms in Figure~\ref{f_m83_line}.  These plots show that the 8/250 and 160/250~$\mu$m ratios trace similar structures offset from the dust mass as well as the H$\alpha$ and 24~$\mu$m emission associated with star formation.  To check how well the 8/250 and 160/250~$\mu$m ratios are correlated, we plot the two ratios in Figure~\ref{f_8-250-160-250}.  The relation has a Pearson correlation coefficient of 0.65, implying that the excitation of PAH emission and the heating of the dust seen at 160~$\mu$m are, to some degree, linked.

In spiral density waves, as mentioned before, large quantities of gas and dust are expected where the ISM is shocked on the upsteam sides of the spiral arms, star forming regions would be found immediately downstream of the shocks, and older stars would be expected further downstream \citep[e.g.][]{1969roberts, 1979elmegreen, 2008tamburro, 2009martinezgarcia, 2011sanchezgil}.  This could cause offset enhancement of PAH emission relative to spiral arm dust lanes, as seen in Figure~\ref{f_m83_line} and as also implied by the relation in Figure~\ref{f_8-250-160-250}, in two possible ways.

One possible explanation is that the dense dust lanes on the upstream edge of the spiral arms severely attenuate the starlight escaping from the photoionising stars within star forming regions, but light easily escapes across the downstream side of the spiral arms where the dust density is lower.  Such a geometrical arrangement of the star forming regions relative to the dust would produce the slight offsets between the 24~$\mu$m emission (tracing obscured star formation) and H$\alpha$ emission (tracing unobscured star formation) seen in most of the profiles in Figure~\ref{f_m83_line} and may also explain the downstream areas with enhanced H$\alpha$ emission in profiles C and F.  If photoionising light is primarily travelling asymmetrically from the star forming regions and if the 160~$\mu$m band traces dust heated by the light escaping from the star forming regions into the diffuse ISM, the 160~$\mu$m emission would appear enhanced relative to 250~$\mu$m emission along the downstream side of the spiral arms.  Similarly, PAHs mixed in with the dust emitting at 160~$\mu$m would be excited by the ultraviolet light escaping from star forming regions and appear enhanced relative to the dust emission in the same locations, although the total PAH emission itself will peak along the spiral arms where the total mass of the PAHs is greater (which is also true for dust emission observed in any single band).

The other possible explanation is that the PAHs and the dust seen at 160~$\mu$m are locally heated by a young, non-ionising population of stars (stars with ages older than 4 Myr) that have left the dusty star forming regions in the spiral arms.  As shown by \citet{1999leitherer}, such a population would still produce a substantial amount of ultraviolet and blue light that could strongly enhance the PAH emission and the temperature of the large dust grains on the downstream side of the arms, but the radiation from these stars may not include higher energy photons that destroy PAHs.  In the profiles in Figure~\ref{f_m83_line}, the ultraviolet emission either peaks downstream of the dust mass or has a profile on the downstream side that is broader than the dust emission profile.  Additionally, the ultraviolet emission appears to stronger relative to the H$\alpha$ emission in most downstream locations.  This provides additional support for the possibility that the 8~$\mu$m emission observed in M83 originates from PAHs excited locally by young, non-ionising stars, although additional analysis would be needed to confirm this.

It is also worth briefly noting that we do not see variations in the 8/160 or 8/250~$\mu$m ratios in M83 that imply a dependence upon radius and hence a dependence upon the metallicity, which decreases with radius \citep{1994zaritsky}.  The variations in the PAH emission with respect to the cold dust emission are driven mainly by local excitation of the PAHs.

\section{Discussion}
\label{s_discussion}

The results show that the 8/24~$\mu$m ratio decreases in many regions with high SSFR, in agreement with previous findings from \citet{2004helou}, \citet{2005calzetti}, \citet{2006bendo}, \citet{2006madden}, \citet{2007berne}, \citet{2007lebouteiller}, \citet{2007povich}, \citet{2008bendo}, \citet{2008gordon}, and \citet{2014calapa}.  Even though our analysis is mainly focused on the 7.7~$\mu$m PAH feature that falls within IRAC channel 4, the results from \citet{2007lebouteiller}, \citet{2007povich}, and \citet{2008gordon} suggest that other PAH emission features may also decrease relative to hot dust emission within star forming regions.

Our results showing the correlation of the 8~$\mu$m emission from PAHs with the far-infrared emission from large dust grains is largely in agreement with the results from \citet{2002haas}, \cite{2008bendo}, and \citet{2014calapa}.  However, the general conclusion from these papers had been that the PAHs are mixed with large grains heated by the diffuse ISRF.  While NGC~2403 certainly fits that scenario, M83 does not.  Instead, PAH emission in M83 is more strongly associated with large dust grains heated either by light escaping from star forming regions and travelling hundreds of pc away from the spiral arms or locally by young stars that produce substantial non-ionising ultraviolet radiation, which was unexpected.  Further study would be needed to determine whether such variations in PAH excitation are seen among other nearby galaxies as well.

PAH may be excited by different radiation fields in NGC 2403 and M83 because of the differences in the spiral structure in the two galaxies.  Because NGC~2403 is a flocculent spiral galaxy, star formation is expected to be triggered by clouds collapsing in local gravitational instabilities \citep[e.g.][]{2003tosaki,2010dobbs}.  In such a scenario, inflowing dust on scales of tens or hundreds of parsecs may be roughly symmetrically distributed around star forming regions, although more modelling work on cloud collapse in flocculent galaxies is needed to confirm this.  If the dust is distributed this way, dust near the centres of these shells would absorb the ultraviolet and blue light from the star forming regions inside, and any PAHs within these central regions would be destroyed.  Meanwhile, dust and PAHs in the outer shells would be shielded from the light from the star forming regions and would instead be heated by the diffuse ISRF.  Also note that the stars in spiral arm filaments in flocculent spiral galaxies are not expected to exhibit any age gradients like grand-design spiral galaxies \citep{2010dobbs}.  If young, non-photoionising stars contribute significantly to PAH excitation, then the relatively homogeneous distribution of these stars may result in the PAHs appearing enhanced over broad areas rather than appearing enhanced near the spiral filaments.  In contrast to NGC~2403, M83 is a grand design spiral galaxy in which cold, dusty gas flows into star forming sites mainly from one side of the spiral arms and star forming regions emerge from the other side \citep[e.g. ][]{1979elmegreen, 1993garciaburillo, 2008tamburro, 2009egusa, 2013vlahakis}.  Hence, dust will preferentially be located upstream of individual star forming regions within M83.  As described in Section~\ref{s_m83pah}, PAHs are destroyed within the centres of these star forming regions, but PAHs in the diffuse ISM could be excited by starlight diffusing out of the optically-thin side of the star forming regions.  Additionally, multiple studies \citep[e.g. ][]{2009martinezgarcia, 2011sanchezgil} have found gradients in the ages of the stellar populations downstream of spiral arms in grand design spiral galaxies.  PAH emission could appear enhanced downstream of star forming regions if the PAHs are destroyed in photoionising regions but strongly excited locally in regions with soft ultraviolet emission from young, non-photoionising stars.  If these geometrical descriptions for the relation between PAHs and excitation sources is accurate, then we should find that PAHs are excited by the diffuse ISRF in other flocculent late-type spiral galaxies while PAHs are excited in regions offset from star forming regions in other grand design spiral galaxies.

Some dust emission models \citep[e.g. ][]{2007draine} and radiative transfer models \citep[e.g. ][]{2011popescu} show PAHs as excited by the radiation fields from all stellar populations regardless of the hardness or intensity of the fields\footnote{The version of the \citet{2007draine} model typically applied to infrared SEDs is usually based on dust heated by a radiation field with the same spectral shape as the local ISRF as specified by \cite{1983mathis}.  When applying the dust model to data, only the amplitude of the radiation field is treated as a free parameter.  However, \citet{2014draine} includes an example of SED fitting with the \citet{2007draine} model in which the spectral shape of the illuminating radiation field is also allowed to vary.}.  Our results show that this approach is an oversimplification of PAH excitation.  New refinements in dust emission and radiative transfer models are needed to replicate how PAHs are excited by radiation fields from different stellar populations within different galaxies and how the PAH/dust mass ratio may change with variations in the hardness of the illuminating radiation field.  For example, \citet{2013crocker} used stellar population synthesis and simplified models of dust and PAH absorption to predict the contributions of different stellar populations to PAH excitation in NGC 628 and found that $\sim$40\% of the PAHs are excited by stars $<10$~Myr in age, $\sim20$\% are excited by stars with ages of 10-100~Myr, and the remainder are excited by stars $>$100~Myr in age.  It is also apparent that PAH excitation changes across spiral density waves (either because of details in the geometry of the star forming regions or because of variations in the stellar populations on either side of the waves), and it would be appropriate to make improvements to radiative transfer models so that they can replicate these effects.

These results have multiple implications for using PAH emission as a proxy for other quantities.  While PAH emission cannot be used on sub-kpc scales to measure accurate star formation rates, groups such as \citet{2008zhu} and \citet{2009kennicutt} have suggested using globally-integrated PAH emission to estimate extinction corrections for optical star formation tracers such as H$\alpha$ emission, thus producing extinction corrected global star formation metrics.  When PAHs are excited by star forming regions, globally-integrated PAH emission should more accurately represent the light attenuated by dust in star forming regions and should provide fairly accurate star formation rates.  When PAHs are excited by the diffuse ISRF, however, the connection between star formation and PAH emission is less clear, and star formation rates calculated using PAH emission could be less reliable.

Previous results showing a relation between PAH emission and far-infrared emission from large dust grains had implied that PAHs could be used as a proxy of dust mass \citep[e.g.][]{2008bendo}.  In cases where the PAHs are associated with dust heated by the diffuse ISRF, this should still be appropriate, although metallicity-related effects would still need to be taken into account.  In cases where the PAHs are heated by diffuse light from star forming regions or from young, non-photoionising stars that have emerged from star forming regions, the PAH emission will still scale approximately with dust mass but will also vary depending upon the radiation field from the young stars.  In this situation, using PAH emission to trace dust mass may be less reliable.

Multiple authors have identified an empirical relation between either radially-averaged or globally-integrated PAH and CO emission \citep{2006regan,2010bendoco, 2013tan, 2013vlahakis}, implying that the PAHs are, to some degree, correlated with molecular gas.  This would be expected if the PAHs also trace the cold dust that is found associated with the molecular gas.  However, the relation between PAH and CO emission breaks down on small spatial scales, including in NGC~2403 (\citealt{2010bendoco}, but also see \citealt{2013tan}).  In M51, \citet{2013vlahakis} found an offset between PAH and CO emission in the spiral arms, with the molecular gas associated with the cold dust in the locations where material is entering the spiral arms and the PAH emission appearing enhanced further downstream where it is excited by light from young stars.  Our results imply that, in future work, we may be able to measure a similar offset between PAH and CO emission in M83 as well as other grand design spiral galaxies.  While PAH emission was already shown to be a poor tracer of molecular gas on sub-kpc scales, the phenomenology of PAHs excitation in spiral density waves causes even more problems with using it as a proxy for molecular gas.

\section{Conclusions}
\label{s_conclusions}

We identified different relations between PAH emission and far-infrared emission from large dust grains in the two galaxies we examined.  For NGC 2403, we find the 8~$\mu$m emission is most strongly associated with emission from cold dust at 250~$\mu$m.  In particular, we find that the 8/250~$\mu$m ratio shows a very strong dependence upon the 3.6~$\mu$m emission from the total stellar population, indicating that the PAHs are mixed in with the diffuse dust and heated by the diffuse ISRF from the total stellar population.  Star forming regions play a much less significant role in the excitation of the PAHs observed in the 8~$\mu$m band. In contrast, we see in M83 that the PAH emission is more strongly associated with the 160~$\mu$m emission from large grains heated by star forming regions as implied by the strong correlation between the 160/250~$\mu$m and 8/250~$\mu$m ratios. This illustrates that PAHs in M83 are excited either by starlight escaping asymmetrically from star forming region so that locations towards the downstream edges of the spiral arms show enhancement in 8~$\mu$m emission compared to the dust mass or that the PAHs are excited locally by young, non-photoionising stars that have migrated downstream from the spiral arms.

Many dust emission and radiative transfer models currently treat PAHs as though they are excited by all radiation fields of all intensities from all stellar populations within the galaxies to which they are applied, much in the same way that emission from silicate and large carbonaceous dust grains is modelled.  The results from just these two galaxies show that this assumption is not universally applicable.  These dust models need to be adjusted to account for the observational results showing that PAHs are sometimes excited by the diffuse ISRF from the total stellar population and sometimes excited either by young, non-photoionising stars that have emerged from star forming regions or light escaping from these regions and travelling hundreds of pc away (although the PAH emission may be inhibited within the star forming regions themselves).  Additionally, some models rely upon using a single SED shape (such as the SED of the local ISRF) for the radiation field illuminating PAHs and dust.  Such models cannot account for the possibility that PAH emission could be inhibited if the radiation fields are excessively hard.  To properly characterise the PAH excitation, it is necessary to model the PAHs as being illuminated by radiation fields with different spectral shapes.

Our data here show differences between the PAH excitation within two galaxies with similar Hubble types.  We should expand the analysis to include galaxies with a wider range of Hubble types, including E, S0, and Sa galaxies where the evolved stellar populations may play a larger role in dust heating and therefore may be more responsible for PAH excitation.  We should also examine other grand-design spiral galaxies to determine whether PAH emission from the spiral arms in these galaxies is offset from the star forming regions in the same way that it is in M83. This research will lead to a better understanding of PAH excitation mechanisms as well as the relation of PAHs to star formation and large dust grains.

\section*{Acknowledgments}

This work has ben produced as part of a MSc Thesis for the University of Manchester. SPIRE has been developed by a consortium of institutes led by Cardiff Univ. (UK) and including: Univ. Lethbridge (Canada); NAOC (China); CEA, LAM (France); IFSI, Univ. Padua (Italy); IAC (Spain); Stockholm Observatory (Sweden); Imperial College London, RAL, UCL-MSSL, UKATC, Univ. Sussex (UK); and Caltech, JPL, NHSC, Univ. Colorado (USA). This development has been supported by national funding agencies: CSA (Canada); NAOC (China); CEA, CNES, CNRS (France); ASI (Italy); MCINN (Spain); SNSB (Sweden); STFC, UKSA (UK); and NASA (USA).  IDL is a postdoctoral researcher of the FWO-Vlaanderen (Belgium).

\bibliography{jonesa_references}

\appendix

\section{Analysis of NGC~2403 using data measured in 27 arcsec bins}
\label{a_binsizecheck}

The two galaxies are located at different distances.  NGC~2403 is at a distance of 3.2~Mpc, while M83 is at 4.5~Mpc; the difference in distances is approximately a factor of 1.5.  However, the analysis relies upon measurements in 18~arcsec bins, which represent different spatial scales in the two galaxies.  To test whether distance effects could have biased our results for NGC~2403, we repeat the analyses on NGC~2403 in Section \ref{s_binned_analysis} and \ref{s_ngc2403pah} using 27~arcsec pixels, which is close to the same spatial scale as the 18~arcsec bins used for the analysis on M83.

\begin{figure}
\begin{center}
\epsfig{file=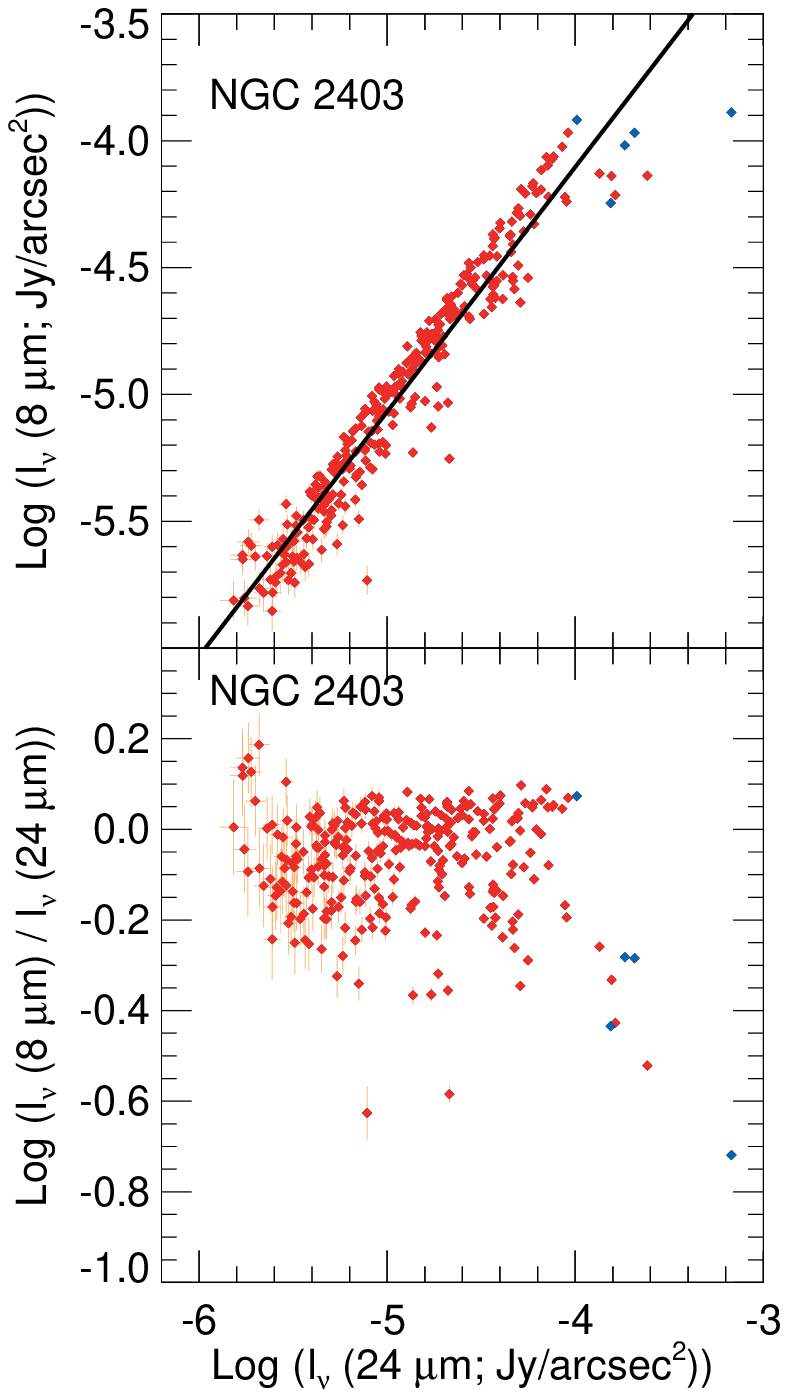}
\caption{The 8~$\mu$m surface brightness and the 8/24~$\mu$m ratios as a function of 24~$\mu$m emission for the 27~arcsec binned data for NGC 2403.  Only data detected at the $5\sigma$ level are displayed.  The best fitting linear functions between the surface brightnesses (weighted by the errors in both quantities) are shown as green lines in the top panels.}
\label{f_8-24_large}
\end{center}
\end{figure}

\begin{figure}
\begin{center}
\epsfig{file=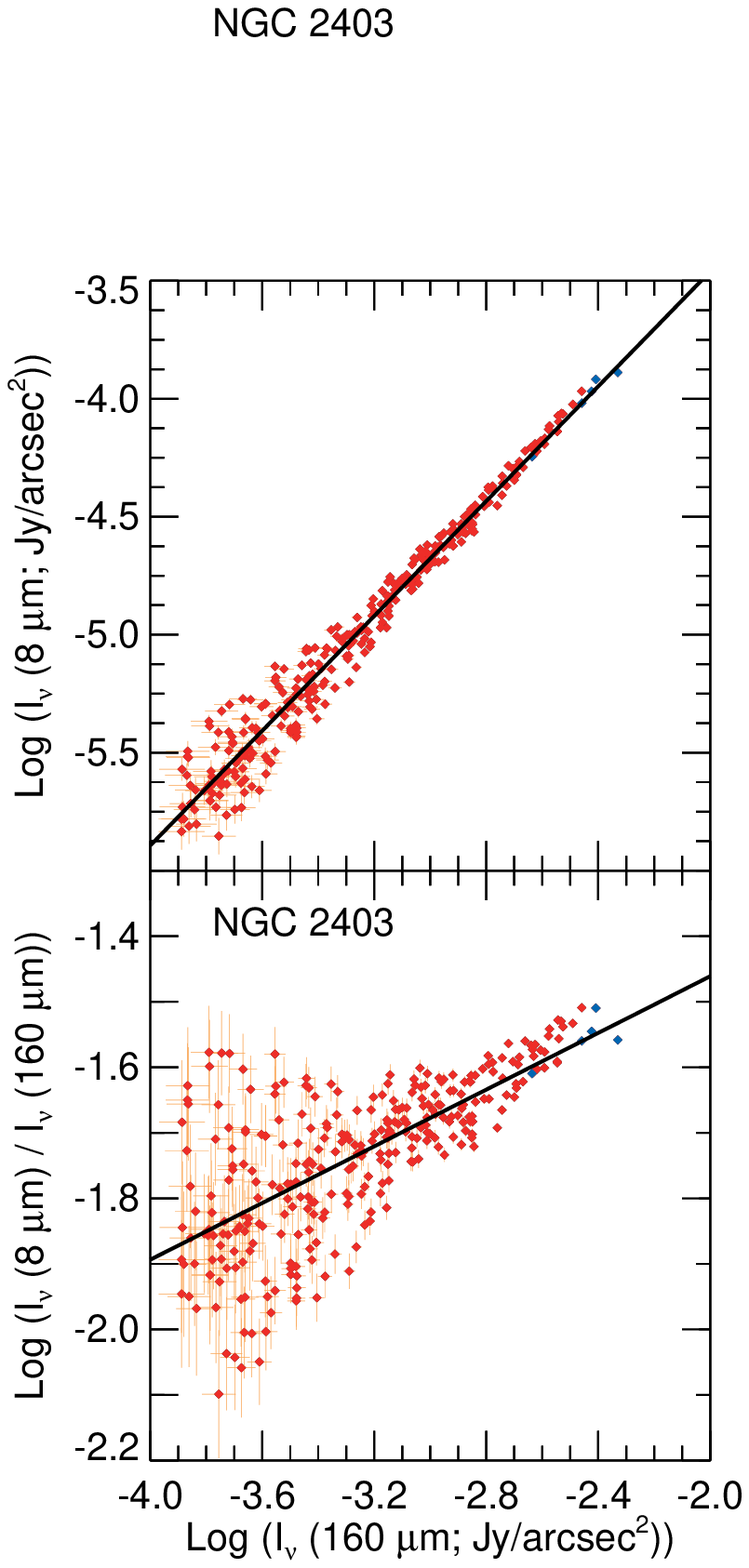}
\caption{The 8~$\mu$m surface brightness and the 8/160~$\mu$m ratios as a function of 160~$\mu$m emission for the 27~arcsec binned data for NGC 2403.  The data are formatted in the same way as in Figure~\ref{f_8-24_large}.}
\label{f_8-160_large}
\end{center}
\end{figure}

\begin{figure}

\epsfig{file=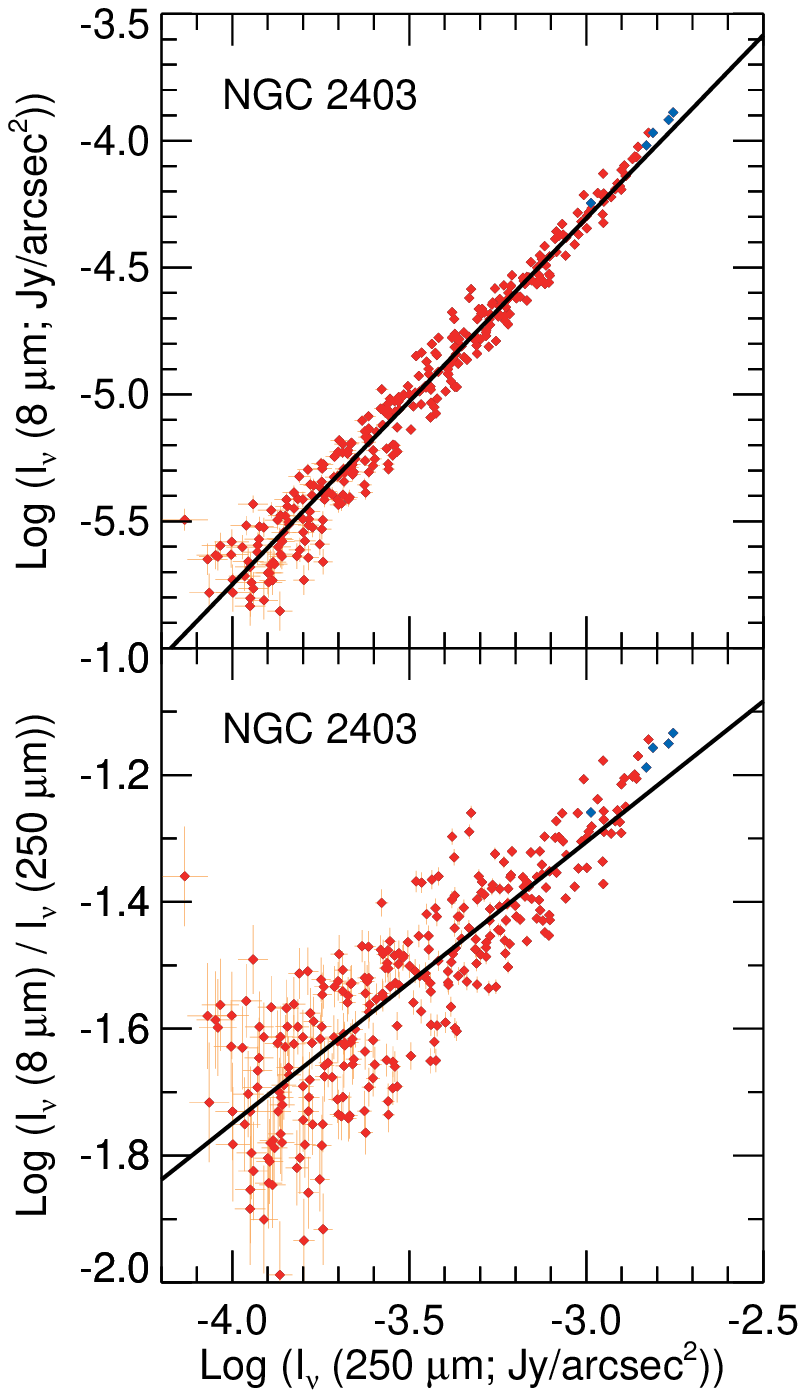}
\caption{The 8~$\mu$m surface brightness and the 8/250~$\mu$m ratios as a function of 250~$\mu$m emission for the 27~arcsec binned data for NGC 2403.  The data are formatted in the same way as in Figure~\ref{f_8-24_large}.}
\label{f_8-250_large}
\end{figure}

For comparison to Figures~\ref{f_8-24}-\ref{f_8-250}, we present in Figure~\ref{f_8-24_large}-\ref{f_8-250_large}, comparisons between the 8~$\mu$m emission and either the 24, 160, or 250~$\mu$m emission as measured within 27~arcsec bins in NGC~2403.  Correlation coefficients for the relations measured in the 18 and 27~arcsec bins are shown in Table \ref{t_correlation_large}.

We see the relations do not change much at all.  The correlation coefficients change by $\leq0.06$, which is negligible.  Given the 27~arcsec binned data, we would reach the same conclusions.  We still see significant scatter in the relation between 8 and 24~$\mu$m data (as exhibited by the low correlation coefficient for the relation between the 8/24~$\mu$m ratio and 24~$\mu$m emission) while we still see very little scatter in the relation between the 8 and 250~$\mu$m emission (as exhibited by the high correlation coefficient for the relation between the 8/250~$\mu$m ratio and 250~$\mu$m emission).  We also still see that the 8/250~$\mu$m ratio is very strongly correlated with the 3.6~$\mu$m emission.  

Ultimately, these results indicate that we can obtain the same results with the coarser resolution data for NGC~2403.  Resolution effects have a negligible effect on our analysis and are therefore not a major concern when comparing the results from NGC~2403 with the results for M83.

\begin{figure}
\begin{center}
\epsfig{file=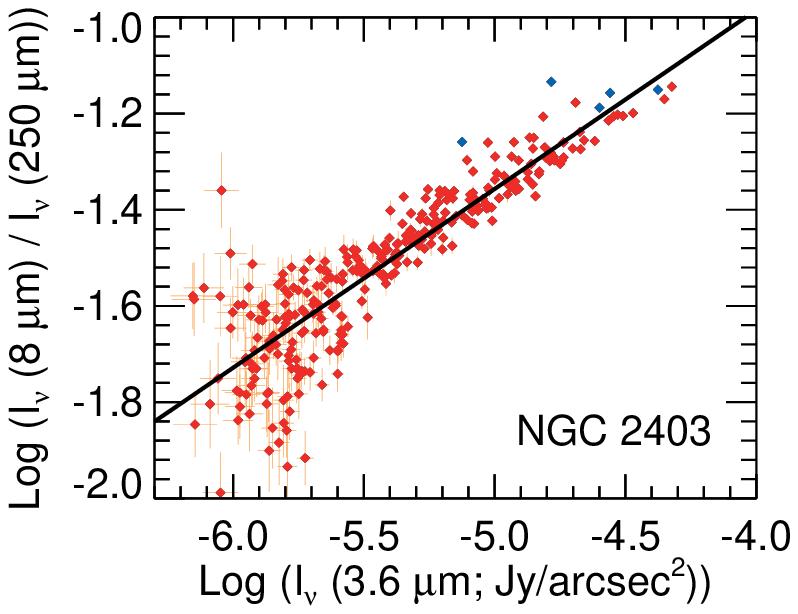}
\epsfig{file=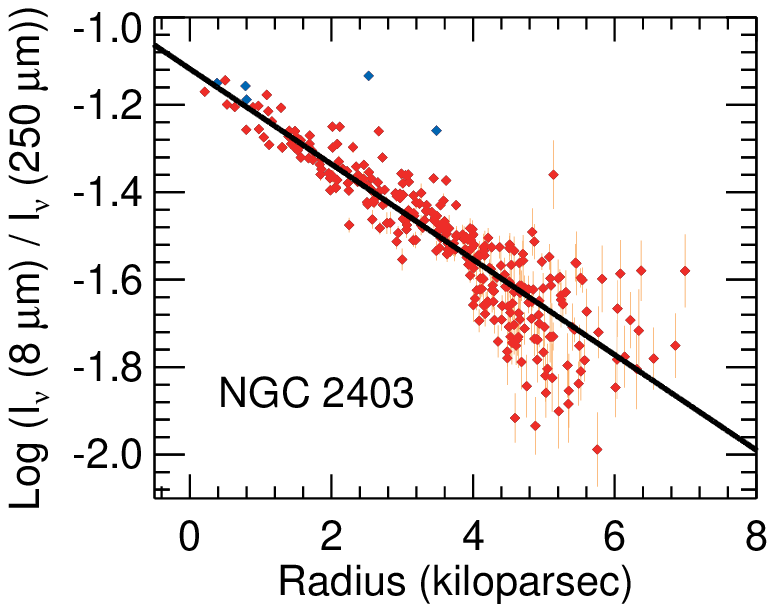}
\caption{The top two panels are the 8/250$~\mu$m surface brightness ratio plotted as a function of the 3.6~$\mu$m surface brightness and galactocentric radius for the 27~arcsec binned data for NGC~2403. The data are formatted in the same way as in Figure~\ref{f_8-24_large}.  The radii are based on using an inclination of $62.9\deg$ from \citet{2008deblok}. }
\label{f_2403-8-250_large}
\end{center}
\end{figure}

\begin{figure}
\begin{center}
\epsfig{file=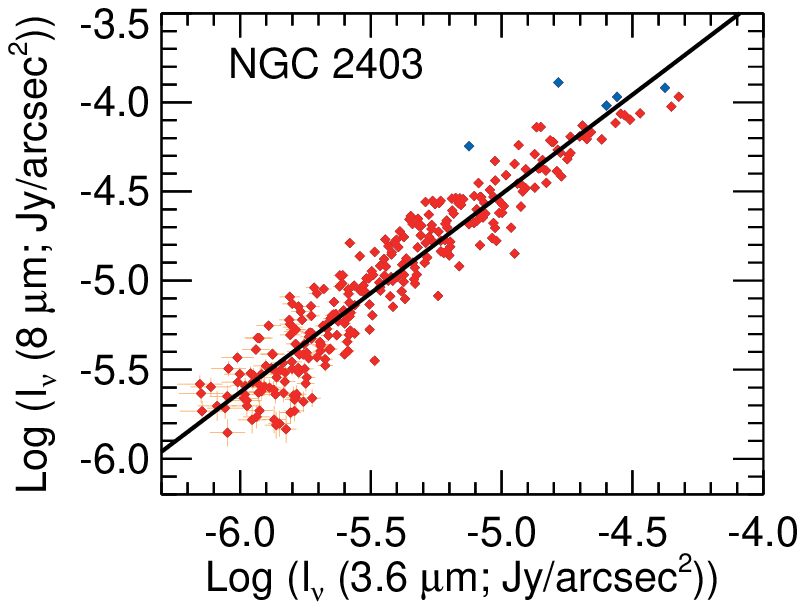}
\epsfig{file=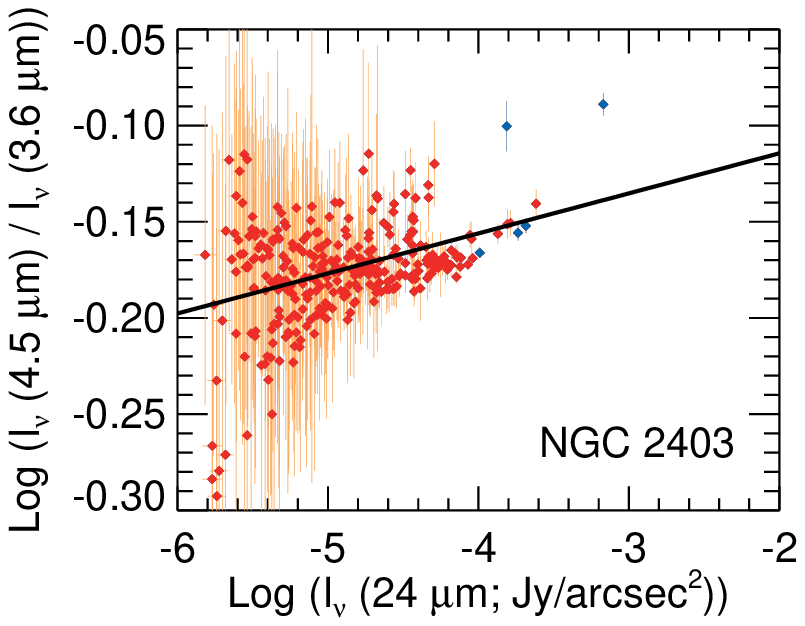}
\caption{The top panel is the 8$~\mu$m to 3.6$~\mu$m relationship. The lower panel plots the 4.5/3.6$~\mu$m ratio vs 24$~\mu$m relationship. The data are formatted in the same way as in Figure~\ref{f_8-24}.}
\label{f_2403-8-3_large}
\end{center}
\end{figure}

\begin{table}
\caption{Pearson correlation coefficients for the binned data.}
\label{t_correlation_large}
\begin{tabular}{p{5.6cm}cc}
\hline
       NGC 2403 & 
   	 &
	\\ 
       	Bin Size (arcsec) & 
   	18 &
	27
	\\ \hline
$\log (I_\nu(8\mu\mbox{m}))$ vs $\log(I_\nu(24\mu\mbox{m}))$ &
   	0.96	&
	0.96	\\
$\log (I_\nu(8\mu\mbox{m})/I_\nu(24\mu\mbox{m}))$ vs $\log(I_\nu(24\mu\mbox{m}))$ &
	-0.13	&
	-0.13	\\
$\log (I_\nu(8\mu\mbox{m}))$ vs $\log(I_\nu(160\mu\mbox{m}))$ &
	0.98	&
	0.98	\\ 
$\log (I_\nu(8\mu\mbox{m})/I_\nu(160\mu\mbox{m}))$ vs $\log(I_\nu(160\mu\mbox{m}))$ &	
	0.65	&
	0.69	\\
$\log (I_\nu(8\mu\mbox{m}))$ vs $\log(I_\nu(250\mu\mbox{m}))$ &
	0.98	&	
	0.98	\\ 
$\log (I_\nu(8\mu\mbox{m})/I_\nu(250\mu\mbox{m}))$ vs $\log(I_\nu(250\mu\mbox{m}))$ &
	0.83	&
	0.84	\\
$\log (I_\nu(8\mu\mbox{m}))$ vs $\log(I_\nu(3.6\mu\mbox{m}))$ &
	0.94	&
	0.95	\\
$\log (I_\nu(8\mu\mbox{m})/I_\nu(250\mu\mbox{m}))$ vs $\log(I_\nu(3.6\mu\mbox{m}))$ &
	0.89	&
	0.88	\\
$\log (I_\nu(8\mu\mbox{m}))$ vs Radius (Kpc) &
	-0.82	&
	-0.88	\\			
	\hline
\end{tabular}	
\end{table}

\label{lastpage}

\end{document}